\documentclass[useAMS,usenatbib]{mn2e}

\usepackage{graphicx}
\usepackage{times}
\usepackage{natbib}
\usepackage{epsfig}

\begin{document}

\title[Cosmic Rays in ENZO.] {Modelling injection and feedback of Cosmic Rays in grid-based cosmological simulations: effects on cluster outskirts.}
\author[F. Vazza, M. Br\"{u}ggen, C.Gheller, G. Brunetti]{F. Vazza$^{1,2}$\thanks{E-mail: f.vazza@jacobs-university.de}, M.Br\"{u}ggen$^{1}$, C.Gheller$^{3}$, G. Brunetti$^{2}$\\
$^{1}$ Jacobs University Bremen, Campus Ring 1, 28759, Bremen, Germany \\
$^{2}$INAF/Istituto di Radioastronomia, via Gobetti 101, I-40129 Bologna,
Italy\\
$^{3}$  Swiss National Supercomputing centre, Via Cantonale, CH-6928 Lugano}

\date{Received / Accepted}
\maketitle
\begin{abstract}
We present a  numerical scheme, implemented in the cosmological adaptive mesh refinement code {\it {\small ENZO}}, to model the injection of Cosmic Ray (CR) particles at shocks, 
their advection and their dynamical feedback on thermal baryonic gas. We give a description of the algorithms  and show their tests against analytical and idealized one-dimensional problems. Our implementation
is able to track the injection of CR energy, the spatial advection of CR energy and its feedback on the thermal gas in run-time. 
This method is applied to study CR acceleration and evolution in cosmological volumes, with both fixed and variable mesh resolution. We compare the properties of galaxy clusters with and without
CRs, for a sample of high-resolution clusters with different dynamical states.   At variance with similar simulations based on Smoothed Particles Hydrodynamics, we report that the inclusion of CR feedback in our method decreases the central gas density in clusters, thus reducing the 
X-ray and Sunyaev-Zeldovich effect from the clusters centre, while enhancing the gas density 
and its related observables near the virial radius.

\end{abstract}

\label{firstpage} 
\begin{keywords}
galaxy: clusters, general -- methods: numerical -- intergalactic medium -- large-scale structure of Universe
\end{keywords}


\section{Introduction}
\label{sec:intro}

Galaxy clusters form through the
hierarchical assembly of dark matter (DM) and gas  over
cosmological times. This process happens via super-sonic
accretion, with efficient driving of turbulent motions, shock
waves and magnetic field compression and amplification in the cosmic gas \citep[e.g.][and references therein]{by00,do08,br11}. 
This picture is based on  observational evidences \citep[e.g.][for recent results]{ct02,gf02,fe08}, and is
theoretically supported by numerical simulations \citep[e.g.][]{ry98,do99,mi00,mi01, ry03, pf07, in08, ho08,ka07, sk08, va09shocks,donn09,va09turbo,va11turbo,bo11}.
However, the details of the interplay between various non-thermal
components of the ICM are still quite uncertain.

Shock waves driven by the accretion of matter constitute the
most important channel for the thermalization of baryonic gas over
cosmological time, and they may also be efficient sites of cosmic ray particles (CR) acceleration.
CRs are thought to be accelerated in the pattern of MHD fluctuations across the shock transition,
in a process known as diffusive shock acceleration, DSA \citep[e.g.][] {be78, bo78, dv81, ebj95, kj90, md01,kj07, ka09, ca10}.
The diffusion of accelerated 
particles across the shock produces a shock-precursor, which modifies the
overall compression produced by the entire shock structure in a non-linear way \cite[e.g.][]{dv81,bl04,ab05}.
The main factors
which govern the overall efficiency of the injection of relativistic particles are the Mach number of the shock, $M$, and
the properties of the magnetic field in the shock transition. At quasi parallel shocks, $M$ is 
the primary parameter which determines the acceleration
efficiency of CRs, while the secondary parameter is the minimum momentum of injection of particles, $p_{\rm th}$, required
to start the acceleration. In the test-particle regime, where the accelerated particles do not have any dynamical feedback on the 
structure of the shock, the DSA predicts the universal energy spectrum of $N(p) dp \propto p^{-q} dp$ for the
accelerated particles, where $q$ is related to the Mach
number only via $q=2 (M^{2}+1)/(M^{2}-1)$. 
However, models of DSA predict that
at strong shocks, $M>5$, the fraction of accelerated CR
particles becomes non-negligible compared to the thermal pool, 
$\xi>10^{-3}$, and that a significant fraction of the shock
energy is transferred to the CR particles, which in turn modify
the shock structure. The result is a larger shock compression and 
an amplification of the magnetic fields in the pre-shock
region \citep[e.g][for a modern review]{ka09}. The basic
predictions of DSA have been tested successfully with multi-band observations of radio and $\gamma$ emission from 
remnants of supernovae \citep[e.g.][]{rey08, vi10, ed11}.


Once accelerated, CR particles can  
accumulate in galaxy clusters \citep{bbp97,volk99} possibly producing an
important non-thermal component which could be directly sampled
by future gamma ray observations \citep[e.g.][] {pi10,bl11b}.
Secondary particles injected in the ICM via proton--proton
collisions may cause detectable synchrotron radiation
\citep[e.g][]{bl99, de00}
and, as well as primary particles, that are 
re-accelerated by MHD turbulence and may thus escape radio halos
\citep{br08}.
CR particles can mix with the thermal plasma, and may have an important dynamical effect
on the ICM \citep[e.g.][]{ro11, bl11}.
Its role can also be that of shaping the evolution of X-ray cavities
powered by AGN jets \citep[e.g.][]{mb07,gu08,ma11}, and to provide
an additional source of thermalization in cool core clusters, via 
Alfv\'{e}n wave dissipation \citep[e.g.][]{gu08}.

\bigskip

The goal of this work is to provide a simplified but robust
algorithm to model the run-time injection  of CR
particles at cosmological shock waves, their advection and their dynamical effect on the
evolution of thermal gas in large-scale structures. 
This 
complements the methods developed by \citet{pf07} and \citet{en07}, where the injection and feedback from CRs is simulated with Lagrangian 
cosmological simulations using smoothed-particle-hydrodynamics. 
To this end, we implemented new modules to follow CRs in the framework of the public
1.5 version of the {\it {\small ENZO}} code \citep{br95}, and
we run a number of validation tests and large-scale production runs in 
cosmology.

The paper is organized as follows: in Sect.\ref{sec:methods} we describe
 the algorithms developed to treat the injection, advection and
dynamical feedback of CRs in {\it {\small ENZO}}; in Sect.\ref{sec:test1} we present
one-dimensional tests for the code and its various options; in Sect.\ref{sec:results} we present our 3--D results in cosmology, obtained
with fixed mesh resolution (Sect.\ref{subsec:fixed}) and adaptive
mesh refinement (Sect.\ref{subsec:amr}). A discussion of the results
is given in Sect.\ref{sec:discussion}, while our conclusion
are summarized in Sect.\ref{sec:conclusions}.

\section{Numerical methods}
\label{sec:methods}
\subsection{ENZO 1.5}

All simulations  in this work were performed using the 
{\it ENZO 1.5} code developed by the Laboratory for Computational
 Astrophysics at the University of California in San Diego 
(http://lca.ucsd.edu). 
  
{\it {\small ENZO}} is an adaptive mesh refinement (AMR) cosmological hybrid 
code highly optimized for high performance computing
\citep{br95, osh04, no07,co11}. It uses a particle-mesh N-body method (PM) to follow
the dynamics of the collision-less Dark Matter (DM) component 
\citep{he88}, and an adaptive mesh method for ideal 
fluid-dynamics \citep{bc89}.

The DM component is coupled to
the baryonic matter (gas) via gravitational forces, calculated from the 
total mass distribution (DM+gas) solving the Poisson equation with 
an FFT based approach.
The gas component is described as a perfect fluid and its dynamics
is calculated by solving conservation equations of mass, energy and momentum over a computational mesh,
using an Eulerian solver based on the 
Piecewise Parabolic 
Method (PPM, \citealt{cw84}). This scheme is 
a higher-order extension of Godunov's shock capturing
method \citep{go76}, and it is at least second--order accurate in space (up
to the fourth--order in 1--D, in the case of smooth flows and small time-steps) and
second--order accurate in time.

For a more detailed overview of the numerical algorithms in {\it {\small ENZO}} and 
of the updated strategies for handling the data-structures in the simulations, we 
refer the readers to the excellent recent review of \citet{co11}.


\begin{figure*}
\includegraphics[width=0.9\textwidth,height=0.4\textheight]{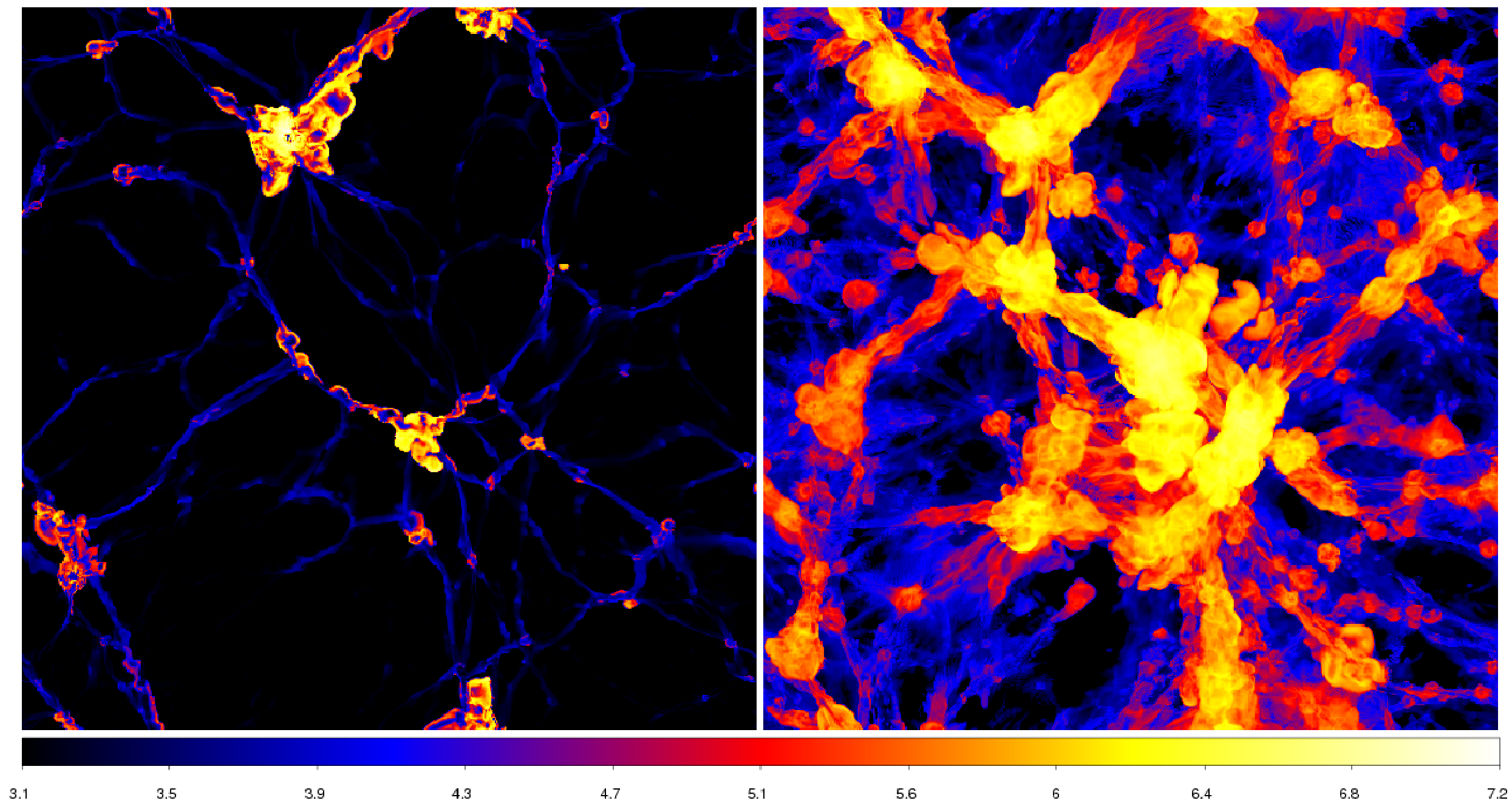}
\includegraphics[width=0.45\textwidth,height=0.4\textheight]{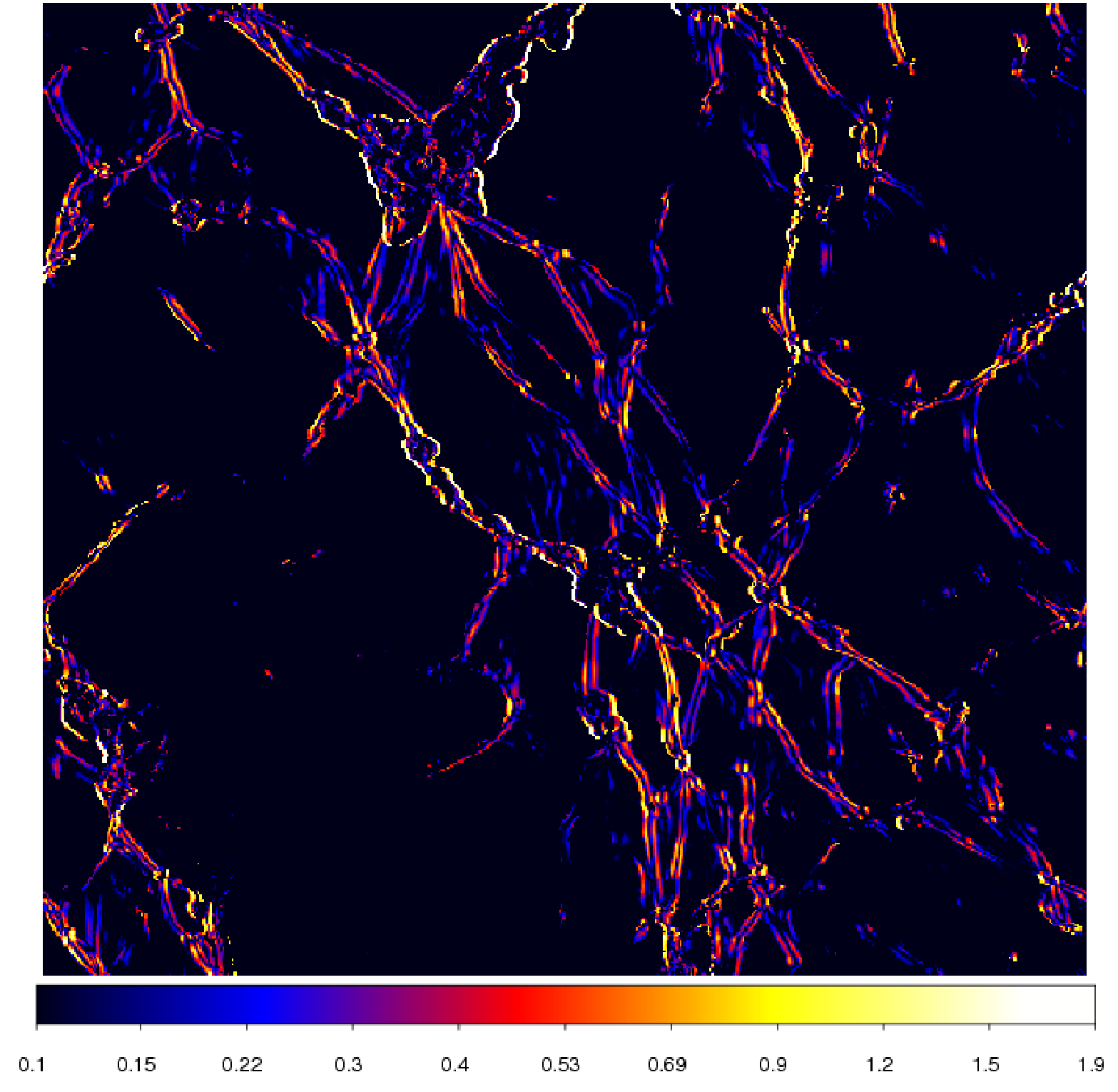}
\includegraphics[width=0.45\textwidth,height=0.4\textheight]{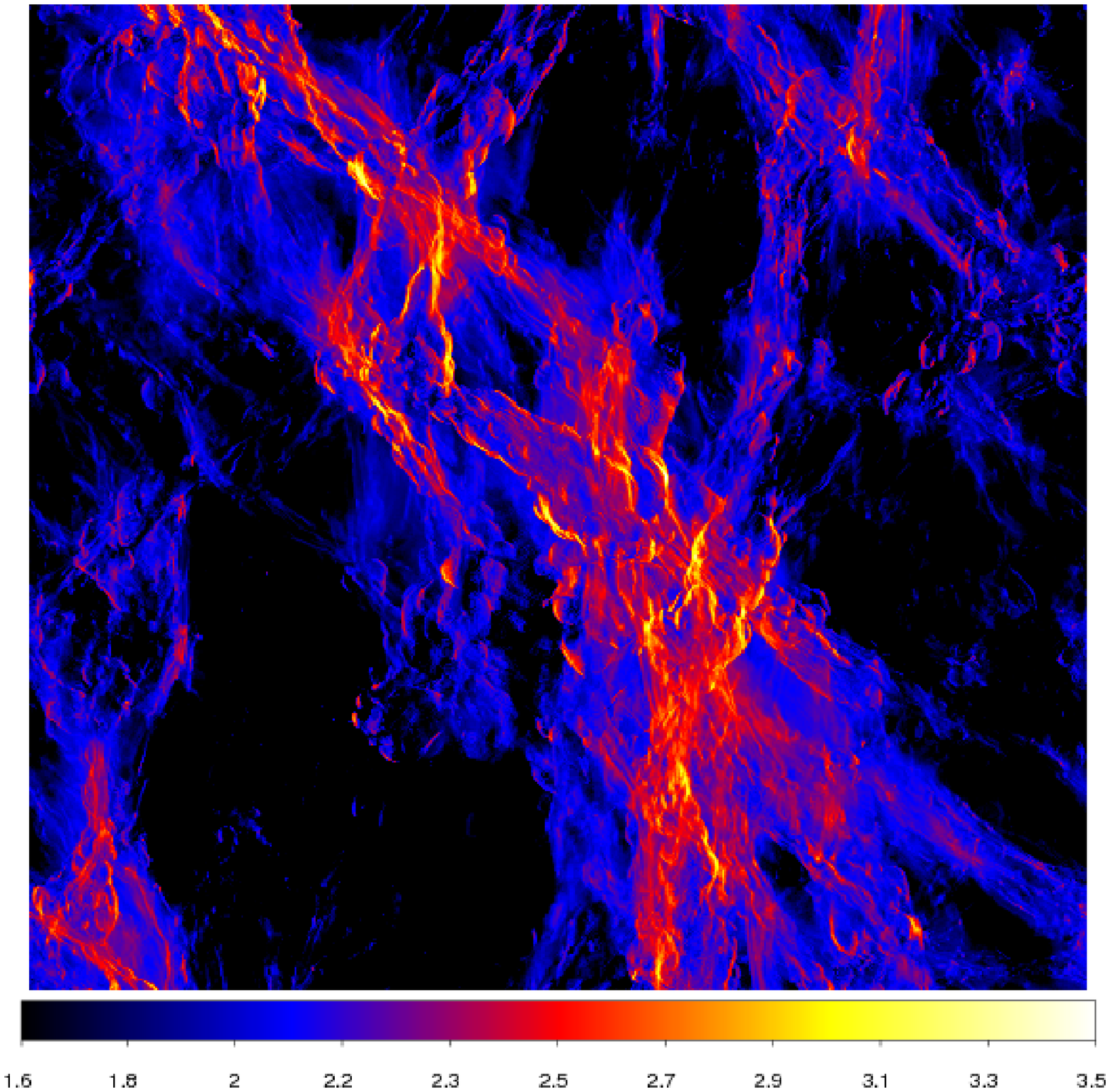}
\caption{2-dimensional slice showing $\log_{\rm 10}(T)$(top panels, in units of [K]) and $\log_{\rm 10}(M)$ reconstructed in run-time by our code (bottom panels), for the full simulated cubic volume with the side of 80 Mpc at $512^{3}$ at $z=0$. The left panels show the 
maps for a slice of depth 300 kpc, while the right panels show
the (volume-weighted) projected maps for a line of sight of 20 Mpc.}
\label{fig:lls1}
\end{figure*}

\subsection{Shock identification}
\label{subsec:shock}

We follow the standard assumption of
Diffusive Shock Acceleration, relating
the injection efficiency of Cosmic Rays at shocks to the 
Mach number of the shock \citep[e.g.][and references therein]{kj07}.

In order to model the injection of CRs at shock waves, a reliable on-the-fly method to measure the Mach number of shocks
in the simulation is required. 

A few grid-based methods in the literature were tested in recent years that apply the Rankine-Hugoniot jumps conditions across shocked cells \citep[e.g.][]{mi00,ry03,sk08,va09shocks}.

In our algorithm, we detect and measure the Mach number of
shock waves during run-time using an approach based on the differences of the gas pressure (or gas plus CR) between cells. 
This method is conceptually very similar to the "Temperature Jump" method of \citep{ry03} and to the "Velocity Jump" method
of \citet{va09shocks}. We decided, however, to use pressure as main
variable of the method, since the CRs pressure is also the physical quantity
used in the code (see Sec.\ref{subsubsec:cr_advect} for a discussion).
  
Our code first selects candidate shocked cells by requiring that 
the flow in the cell is converging: $\nabla \cdot {\bf v} < 0$.
Then the 3--D distribution of cells around the candidate cell is analysed
with one-dimensional scans along the three axes, and it is checked that the gas temperature, $T$, and the gas pseudo-entropy, $S=P \cdot \rho^{-\gamma}$ (where $\rho$ is the gas density and $\gamma$ is the adiabatic index) change in
the same direction, $\nabla S \cdot \nabla T>0$ \citep[e.g.][]{ry03}. 

The gradient of temperature then sets the candidate post-shock and pre-shock
cells.
The Mach number is then evaluated from the information 
of the pressure jump between cells given by the Rankine-Hugoniot condition:

\begin{equation}
\label{eq:mach_press}
\frac{P_{\rm 2}}{P_{\rm 1}}=\frac{2 \gamma \it M^{2}-\gamma+1}{\gamma+1},
\end{equation}

where $P_{\rm 2}$ ($P_{\rm 1}$) is the gas pressure where the temperature across the 
candidate cells is the highest (the smallest). By construction, we consider pre-shock and post-shock cells with a stencil of 3 cells around the minimum of the 3--D divergence.

\begin{figure*}
\includegraphics[width=0.99\textwidth]{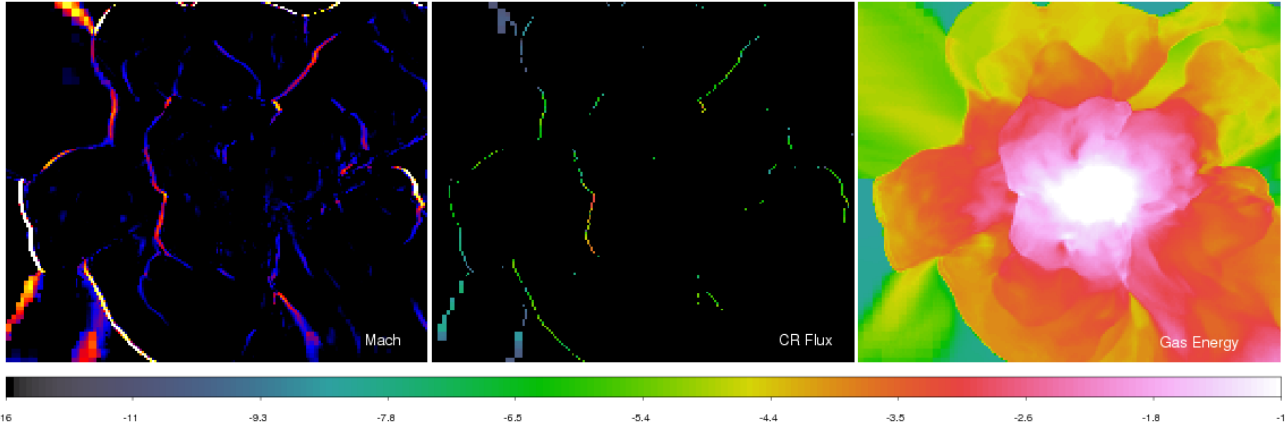}
\caption{2-dimensional slices showing the Mach number (left, color coding
as in the bottom left panel of Fig.\ref{fig:lls1}), the 
energy injected into CR during the last time step (central panel, in ${\rm log_{\rm 10}}[code units]$) and gas energy (same color coding of central panel) for a 
slice with side of  $\sim 5$ Mpc along a line of sight of 32 kpc. 
The simulation employs nested initial conditions and 5 levels of mesh refinement. The fact that only cells with $M>2$ are used for the injection of CRs, explain the small difference in morphologies between the left and the central panel in the
Figure (see Sec.\ref{subsubsec:CR} for details).}

\label{fig:lls2}
\end{figure*}

In the case of candidate shocked cells with several Mach numbers along the different axes, and in case of multiple candidate shocked cells along the same direction, we "clean" the preliminary map of Mach numbers by retaining only the 
maximum Mach number provided by the algorithm, for the considered patch of 
cells \citep[e.g.][]{ry03}. 
Obliquity in shocks can also lead to a small additional smoothing of the shock transition across a few cells \citep[e.g.][]{tasker08,sk08}. We verified that in the case of oblique
shocks in 3--D, the Mach number reconstructed by our procedure is equal, within a $\sim 5-10$ percent even for strong shocks, to that measured for perpendicular shocks having the same input Mach number. Given our cleaning procedure on multiple shocked cells along the same scan direction, our code also makes sure than the reconstructed pattern of shocks, which is used for the injection
of CR energy,  is always 1 cell thick (see the first two maps of Fig.\ref{fig:lls2}).

\begin{figure}
\includegraphics[width=0.45\textwidth]{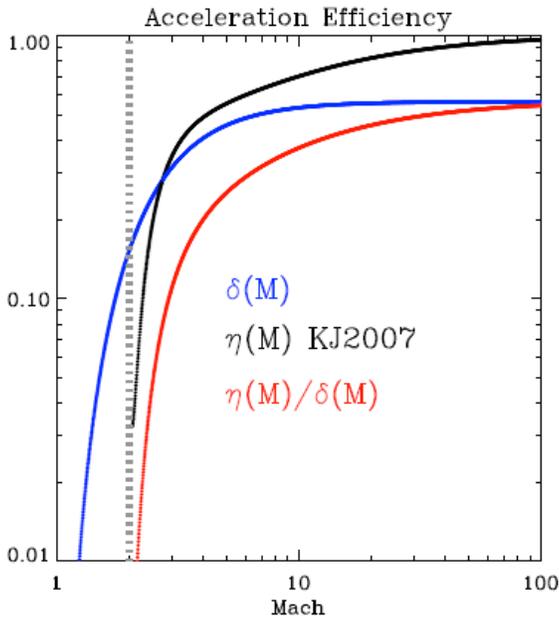}
\caption{Trend with the Mach number of the acceleration efficiency
at shocks for the \citep{kj07} efficiency ($\eta(M)$, in red),
for the standard thermalization efficiency in the post-shock ($\delta(M)$, in blue) and for the ratio of the two (black). In this plot, the kinetic energy flux across the surface of the shock is normalized to one.}
\label{fig:eta}
\end{figure}

Based on our tests \citep[][]{va09shocks}, the use of a
stencil of $\sim 3$ cells ensures in general the best
reconstruction of the shock jumps in cosmological runs. Also, this avoids the 
contamination of density/temperature fluctuations associated to substructures in clusters (e.g. small gas clumps) when the Mach number is computed. 
Usually, only very strong shocks ($M \gg 10$)
are spread across more than 3 cells in cosmological simulations, so only
in these cases we underestimate $M$ slightly. However, these shocks are very rare and not relevant
for the production of CR energy in clusters \citep[e.g.][]{ry03,pf06,sk08,va09shocks}. Given the shape of the
acceleration efficiency we use here (see Sect.\ref{subsubsec:CR} and Fig.\ref{fig:eta}),
the exact value of shock strength is not crucial for $M>20-30$, since the
assumed efficiency function saturates for very strong shocks. 

\bigskip

Figure \ref{fig:lls1} shows the results of our run-time detection scheme for a cosmological box of $80$ Mpc simulated with a fixed grid resolution (Run1\_h in Tab.\ref{tab:tab1}, $\Delta x = 156$ kpc).  We show volume-weighted maps of gas temperature (top panels) and of Mach number reconstructed at run-time by our algorithm (bottom panels) for a thin slice of 300 kpc (left panels) and for a much larger slice with the thickness of 20 Mpc (right panels).
Similar to what is found in other papers \citep[e.g.][]{mi01,ry03,sk08,va09shocks}, the outer regions of clusters and filaments are surrounded by strong, $M \gg 10$,
regular patterns of shocks, while the hotter innermost regions are crossed by more irregular and weaker shocks. 
Even when seen in projection along a large simulated volume (right panels), our reconstructed shocks trace sharp surfaces surrounding 
galaxy clusters and filaments, representing efficient and well-confined sites for the injection of CRs inside large-scale structures. We remark that this is already at variance with cosmological SPH simulations, where the outer shocks are much smoother due to the large smoothing length outside large-scale structures
\citep[][]{pf07,ho08,va11comparison}.

In previous work, we demonstrated that an algorithm based on
velocity jumps (VJ) between cells may present small advantages in the reconstruction of shock waves outside of clusters, owing to a smaller
scatter associated with velocity-based measurements compared to temperature-based
measurements \citep{va09shocks}. However, the vectorial analysis of the 3--D velocity field and the "cleaning" algorithms associated with the procedure, requires a substantially greater 
amount of calculations compared to scalar-based methods. This
is not an issue for any post-processing analysis of shock waves,
but it makes a significant difference in performance when the computation is performed in run-time
(due to 3--D loops and hig her memory usage). For this reason, we considered a pressure-based method
a good compromise for the run-time use needed here.

\begin{figure}
\includegraphics[width=0.46\textwidth]{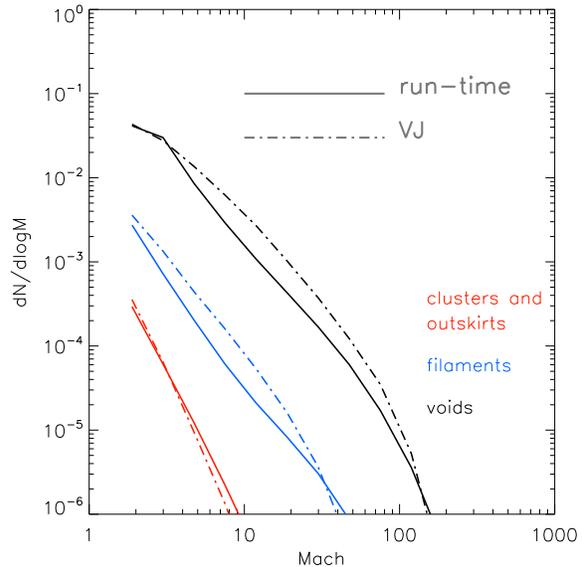}
\caption{Volume distribution of Mach numbers in a simulated cosmological
box of side 80 Mpc (Run1\_h, see Table \ref{tab:tab1}). The solid 
lines report the results for the run-time shock finder employed in
this work, the dot-dashed lines are for the velocity jumps method
introduced in \citet{va09shocks}, run over the
same data. The different colours show the results for different
cosmic environments.}
\label{fig:mach_histo}
\end{figure}

In Fig.\ref{fig:mach_histo} we show the differential
volume distribution of Mach numbers for the full simulated cosmological volume, measured with the two methods.
The distributions of shocks are divided according to the cosmic environment, depending on the total matter over density
of shocked cells, $\delta_{\rm \rho} \equiv \rho_{\rm tot}/\rho_{\rm cr}$ (where 
$\rho_{\rm}$ is the total gas+DM density and $\rho_{\rm cr} \approx 9.31 \cdot 10^{-30} {\rm g/cm^{3}}$ is the critical density of the Universe:
voids, $\delta_{\rm \rho}<3$, filaments, $3 \leq \delta_{\rm \rho} < 30$,  and galaxy clusters and their outskirts, $\delta_{\rm \rho} \geq 30$,
as in \citealt{va09shocks}).

The distributions are very similar to what has been found in our previous work based on
fixed resolution simulations in {\it {\small ENZO}} \citep[e.g.][]{va11comparison}. At the scale of galaxy clusters outskirts and galaxy clusters virial volumes ($\delta_{\rm \rho} > 30$) the two methods yield nearly identical results. 

\bigskip
In runs where adaptive mesh refinement (AMR) is turned on, the Mach number is 
computed at all refinement levels, using a stencil of cells at the same level.
The Mach number estimated at the highest resolution is always preferred in the case of several
Mach numbers for the same cells, obtained with different AMR levels. 
In this paper, we employ adaptive mesh refinement using
the criteria outlined in \citet{va09turbo}. This
means that an AMR
criterion based on velocity jumps in 1--D is added to the usual
criteria of gas and DM over density (e.g. O'Shea et al. 2004), 
ensuring that virtually {\it all} shocks with ($M>2$)
within the AMR regions are re-sampled
down to the maximum available resolution. Hence, almost the entire
population of shocks contributing to CR acceleration (Sect.\ref{subsubsec:CR})
is reconstructed and measured using the highest available resolution in
the domain.
Figure \ref{fig:lls2} shows the pattern of Mach numbers reconstructed in run-time by our method (left panel) for a galaxy cluster.
It also shows
the energy flux of CRs and the gas energy for a runs employing 5 levels of mesh refinement 
(the slice has a side length of $\sim 5$ Mpc and a depth along the line of sight of $32$ kpc). 

In the case of a mixture of gas and CRs, the total
Mach number of the shocks can be obtained by Eq.\ref{eq:mach_press},
provided that the total pressure and an effective adiabatic index
(as $\gamma_{\rm eff}=(\gamma P_{\rm g}+ \gamma_{\rm cr} P_{\rm cr})/(P_{\rm g}+P_{\rm cr})$, \citealt{en07}, where $\gamma_{\rm cr}=4/3$).

\subsection{Method for Cosmic Rays}
\label{subsec:methods}

In the following sections we describe our numerical implementation of the injection, the spatial advection and the pressure feedback of CRs.

In this work, we consider only CR {\it protons}, for which to first approximation radiative and Coulomb losses and the inelastic proton-proton
collisions can be neglected, for the time-scales of interest here \citep[e.g.][]{mi07}. The extension to CR electrons and the modelling of radiative, Coulomb losses and proton-proton collisions
for both species will be investigated in future work.
Also, we only consider CRs generated at large-scale structures shocks.  The inclusion of other sources of CRs, such as supernovae or AGN, as well as the 
coupling of our method for CRs to the new MHD version of {\it {\small ENZO}} \citep{co11} will also
be the subject of future work.

In summary, in our implementation we inject the energy of CR particles at shocks during run-time assuming Diffusive Shock Acceleration (\ref{subsubsec:CR}), we remove this energy from the thermal gas energy pool (\ref{subsubsec:reduced}), we advect it in the simulated volume along with the thermal gas energy (\ref{subsubsec:cr_advect}), and we
compute its pressure
feedback by adding to the total pressure (\ref{subsubsec:cr_feedback}). In order to preserve numerical stability, new time stepping criteria are
added to the standard ones (Sec.\ref{subsubsec:time_step}).

\subsubsection{The injection of Cosmic Rays}
\label{subsubsec:CR}

In our treatment of DSA, we rely on the results of numerical, mostly one-dimensional studies 
of CR acceleration at shock waves as a function
of background plasma parameters  \citep[e.g.][] {md01,kj07, ka09, ca10}

The injection of CR energy in 1--D quasi parallel shocks 
has been studied in detail with advection-diffusion equations
for a variety of shocks, for velocities in the range $150 \leq v_{\rm s} \leq 4500$ km/s and for pre-shock temperatures of $10^{4}-10^{6}$ K. 
In these models the partial dissipation of shock kinetic energy into the
excitation of Alf\'{v}en modes is accounted for, while in our treatment we neglect the presence of pre-existing CR
energy in computing the injection of new CR energy. This is clearly
a limitation of our treatment. However, self-consistent recipes to compute
the estimated acceleration efficiency as a function of both the shock
Mach number and the ratio of CR energy to gas thermal energy at every
shock are presently unavailable, since they require the full self-consistent solution
of diffusion-convection equations \citep[e.g.][]{kj07,ca10}. 

Based on these theoretical papers, we assume that a small fraction of the incoming protons is instantly
accelerated at the dissipative gas sub-shock to a speed larger than the
post-shock sound speed, and that it contributes to the CR pressure inside the cell.

For every detected shock we compute the co-moving shock speed $v_{\rm s}=M c_{\rm s}$ and the resulting energy flux of accelerated CR protons, as a function of the Mach number:

\begin{equation}
\label{eq:fcr}
\phi_{\rm cr} = \eta (M) \cdot \frac{\rho_{\rm u} v_{s}^{3}}{2}=\eta (M) \cdot \frac{\rho_{\rm u} c_{s}^{3}M^{3}}{2} ,
\end{equation}

where $\rho_{\rm u}$ is the co-moving up-stream gas density and $\eta(M)$ is a function of $M$, whose numerical approximation can be found in \citet{ka07}. 
For simplicity we restrict the acceleration of CRs to $M \geq 2$, given that 
the acceleration of CRs for very weak shocks is expected to be extremely small, and it is still very poorly constrained \citep{kr10}.
The trend with Mach number of $\eta(M)$ in the 
\citet{kj07} model is shown
in Fig.\ref{fig:eta}, along with the trend of the
standard thermalization efficiency from Rankine-Hugoniot
jump conditions ($\delta(M)$) and the ratio between the
two. 

To obtain the injected energy density of CRs within each cell, 
we integrate the CR energy flux over the time step:

\begin{equation} 
\label{eq:Ecr}
e_{\rm cr}= \frac{\phi_{\rm cr}}{\rho} \cdot \frac{dt_{\rm l}}{dx_{\rm l}} ,
\end{equation}
 
where $\rho$ is gas density in the cell, $d x_{\rm l}$ is the cell size at the refinement level $\rm l$ and $dt_{\rm l}$ is the time step of the simulation at that refinement level.
The CR energy density, $e_{\rm cr}$, is therefore our main variable to simulate the evolution and dynamics of relativistic particles in the two-fluid model, and it represents the integrated energy density over the energy/momentum spectra
of the CR distribution, $e_{\rm cr} \propto \int{E \cdot N(p) dp} \propto \int{p^{4} \cdot f(p) \cdot (1+p^{2})^{-1/2}dp}$. 

In every cell, the contributions of the CR energy fluxes from
all directions are
 added, and the CR energy at the various AMR level is handled as all other primary quantities in {\it {\small ENZO}}
\citep[see][for a review]{co11}.

\subsubsection{Reduced thermalization at shocks}
\label{subsubsec:reduced}

The transfer of energy from the thermal gas
to CRs causes a reduced thermalization efficiency in the post-shock region \citep[e.g.][]{ab05}. 
This effect can be modelled by removing the corresponding energy flux of accelerated CR from the post-shock region, thus ensuring energy conservation 
where Eq.\ref{eq:Ecr} is applied. 

In {\it {\small ENZO}}, the ''dual energy formalism'' \citep[][]{ry93,br95} is used, which means that whenever the internal gas energy, $e_{\rm g}$, cannot be
calculated correctly as the difference between total, $e_{\rm tot}=e_{\rm g} + v^{2}/2$ (where $v$ is the modulus of the velocity field), and kinetic energy (due to the fact that $e_{\rm tot}\gg e_{\rm g}$ in highly supersonic flows, and that there round-off errors dominate) the total energy is evolved.
Since the acceleration of CR affects the thermalization efficiency of post-shock regions, we need to update the two energies in a consistent way.
Once $e_{\rm cr}$ has been computed (Eq.\ref{eq:Ecr}), we update the values of the total energy and 
of the gas internal energy within the cell according to: $e_{\rm g}'= e_{\rm g} - e_{\rm cr}$ and  $e_{\rm tot}'=e_{\rm tot} - e_{\rm cr}$. This step is performed before starting the Riemann solver of the PPM method at the following time step. 
In cosmological runs without heating from the re-ionization
field (e.g. \citealt{hm96}), the temperature of
the background gas can reach un-realistically low temperatures, $T < 10$ K, with extremely strong shocks ($M>1000$, \citealt{va09shocks}).  In these cases, one may end up in the 
situation of having 
$e_{\rm g}-e_{\rm cr}<0$ in some cells, due to similar round-off errors as in the estimate of $e_{\rm g}$.
Considering that this problem arises only in extremely rarefied regions and in the absence of a
re-ionization background, we just fixed it by forcing our version of {\it {\small ENZO}} not 
to inject new CRs if this is going to 
produce negative values of $e_{\rm g}$ or $e_{\rm tot}$. This fix has no consequence on all thermal
and non-thermal statistics of our simulated large-scale structures, and this problem does not arise when
realistic models for background radiation from re-ionization are adopted.

\begin{figure*}
\includegraphics[width=0.975\textwidth]{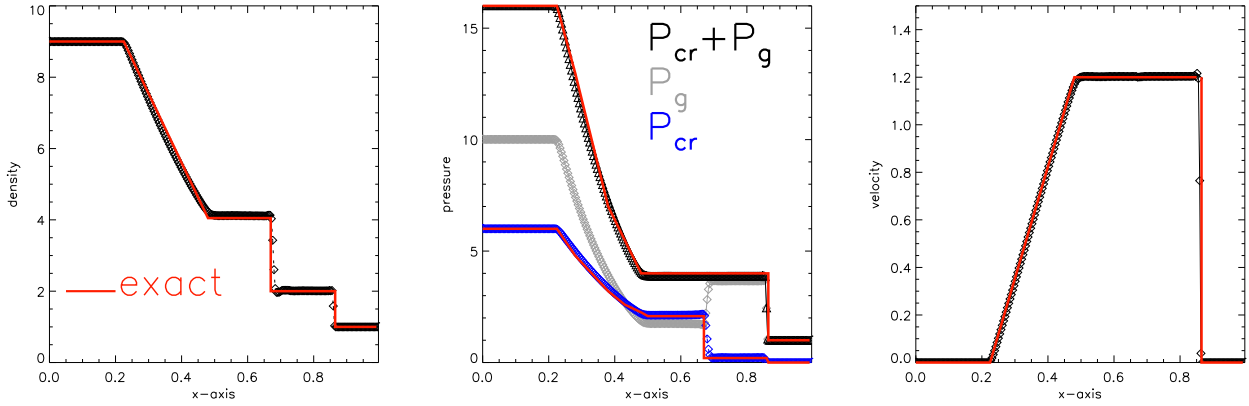}
\caption{Shock tube test for $P_{\rm cr,l}=6$, $P_{\rm g,l}=10$
and  $\rho_{\rm l}=9$, and assuming an acceleration efficiency of $\eta \approx 0.05$ at the shock front. 
The left panel shows the profile of gas density, the
central one the profiles of $P_{\rm g}$, $P_{\rm cr}$ and $P_{\rm tot}$, while
the right one shows the profile of velocity. The diamonds represent the
solution for our code, while the solid red lines shows the 
numerical solution obtained by \citet{mi07}.}
\label{fig:tube1}
\end{figure*}

\subsubsection{Spatial advection of CR energy}
\label{subsubsec:cr_advect}

Once injected, the CRs are spatially advected 
assuming they are "frozen" into the gas velocity field. 
This follows from neglecting the effect of the spatial diffusion
of CRs during the simulation.

Run-time processing of spatial diffusion of CRs is 
computationally expensive, and requires the knowledge of the $\vec{B}$
field topology and of MHD modes in the simulated volume. 
However, according to quasi-linear theory, the spatial diffusion is not expected to play a role
for the physical scales we study here, $L>25-100$ kpc, given that
the typical diffusion time of $\sim$ GeV particles in the 
ICM is of the order of $\sim 1-10$ Gyr for scales of $L \sim 100$ kpc 
assuming $B \sim 0.1-1 \mu G$ \citep [e.g.][as a recent review]{blasi07}.
Since the bulk of CR energy is injected at $z \leq 1$  (Sect.\ref{subsec:fixed}),  the effect of 
spatial diffusion is small compared to  the typical scales analysed in our 
simulations, $\sim 100-1000$ kpc. However, the distributions
of CRs can be slightly modified by the action of diffusion, which may smooth
our distributions on $L_{\rm diff} \sim 50-100$ kpc. Since the self-consistent inclusion 
of diffusion (and conduction) processes in the simulated ICM is still challenging \citep[e.g.][]{ruszkowski11},
and since a self-consistent magnetic field is not included in our version of {\it {\small ENZO}}, we consider
the above as an unavoidable limitation of our present method.

No other dynamical exchange between CRs and thermal gas is assumed to take place (e.g. Coulomb losses), and the coupling between CRs and baryonic gas proceeds via
the momentum equation (see below).

The energy of CRs is advected in space by solving:
\begin{equation}
\frac{\partial E_{\rm cr}}{\partial t}+ \nabla \cdot (E_{\rm cr} {\bf v}) + P_{\rm cr} \cdot \nabla \cdot {\bf v}=0 ,
\label{eq:CR_advect}
\end{equation}

\noindent where $E_{\rm cr}=\rho e_{\rm cr}$, and where we assume a {\it constant} pseudo-adiabatic index for the CR energy (e.g. Kang \& Jones 1990; Mathews \& Guo 2011). Consequently, cosmic ray pressure, $P_{\rm cr}$, and energy , $e_{\rm cr}$, are related via

\begin{equation}
P_{\rm cr}=(\gamma_{\rm cr} -1)\rho e_{\rm cr}=(\gamma_{\rm cr}-1) E_{\rm cr}  \ .
\label{eq:pcr}
\end{equation}

Equation \ref{eq:CR_advect} is numerically solved as:

\begin{eqnarray*}
\label{eq:CR_advect_num}
E_{\rm cr,i}^{n+1}=E_{\rm cr,i}^{n}+dt/dx[(E^{n+1/2}_{\rm cr,i-1/2} \cdot v^{n+1/2}_{i-1/2} 
\nonumber\\
-E^{n+1/2}_{\rm cr,i+1/2}\cdot v^{n+1/2}_{i+1/2} )
+ \,P_{\rm cr}^{*}\cdot(v^{n+1/2}_{i-1/2}-v^{n+1/2}_{i+1/2})],
\end{eqnarray*}

\noindent where

\begin{equation}
 P_{\rm cr}^{*}=\frac{\gamma_{\rm cr}-1} {2}  \cdot (E^{n}_{\rm cr,i-1/2}+  E^{n}_{\rm cr,i+1/2}  ) 
\end{equation}

\noindent is the ''centred''  CR pressure between the the two faces of the cell
 (e.g. Bryan et al. 1995).

Equation \ref{eq:CR_advect_num} is computed
similarly to the total energy and internal energy equations in {\it {\small ENZO}},  via one-dimensional 
sweeps along each direction and using the information of the
fluxes, $v^{n}_{i-1/2}$ and $v^{n}_{i+1/2}$, reconstructed by the 
PPM method (e.g. Bryan et al. 1995).

The effects of cosmological expansion on CRs is treated separately, 
by updating the CR energy density, $e_{\rm cr}$:

\begin{equation}
\frac{\partial e_{\rm cr}}{\partial t}=-3\frac{(\gamma_{\rm cr}-1) \dot{a}}{a} e_{\rm cr},
\end{equation}

\noindent where $a$ and $\dot{a}$ are the cosmological scale factor, and  the scale factor rate of change at each time step (e.g. Bryan et al. 1995).

The real value of the CR adiabatic index

\begin{equation}
\gamma^{*}_{\rm cr}=|\frac{d\log P_{\rm cr}}{d\log \rho}|
\end{equation}

cannot be derived self-consistently, since this would require the knowledge of
the distribution of the CR momenta, $f(p)$.
Depending on the shape
of the momentum distribution and of the minimum momenta, $q_{\rm min}$, this may range from $\gamma^{*}_{\rm cr}=4/3$ to $\gamma^{*}_{\rm cr}=5/3$ (e.g. Ensslin et al. 2007). 
In numerical simulations of DSA, where the diffusion-convection equation
for the evolution of the $f(p)$ is solved in run-time, and coupled 
to the dynamical equation of the gas evolution (e.g. Kang \& Jones 2002), the adiabatic index can be computed self-consistently.
In our case, the value of $\gamma^{*}_{\rm cr}=\gamma_{\rm cr}$ can only be assumed a-priori.
In the following, we will produce tests and simulations where we  fix the adiabatic index to its most extreme value of 
$\gamma_{\rm cr}=4/3$.

\begin{figure*}
\includegraphics[width=0.95\textwidth]{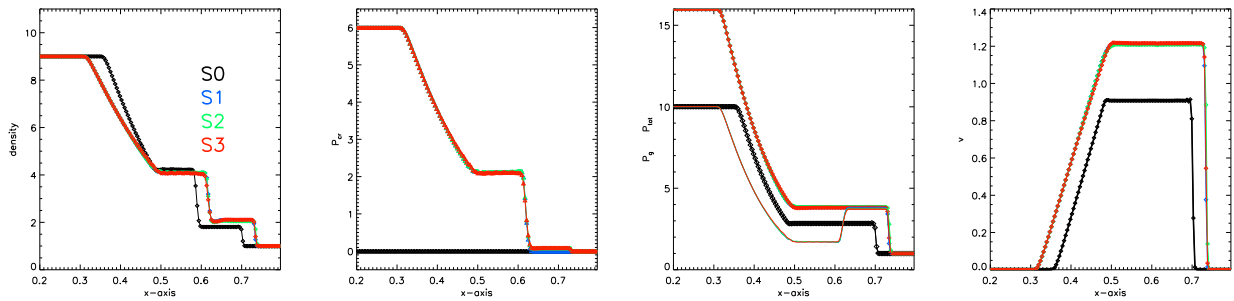}
\caption{Shock tube test for the various physical modules
available to our method. From left to right, we show: gas density, CR pressure, total (thick lines) and gas (thin lines) pressure,
velocity field. The initial conditions are as
in Fig.\ref{fig:tube1}, and $\eta=0.02$ is assumed at
the shock front. }
\label{fig:tube2}
\end{figure*}

\begin{figure}
\includegraphics[width=0.45\textwidth,height=0.4\textwidth]{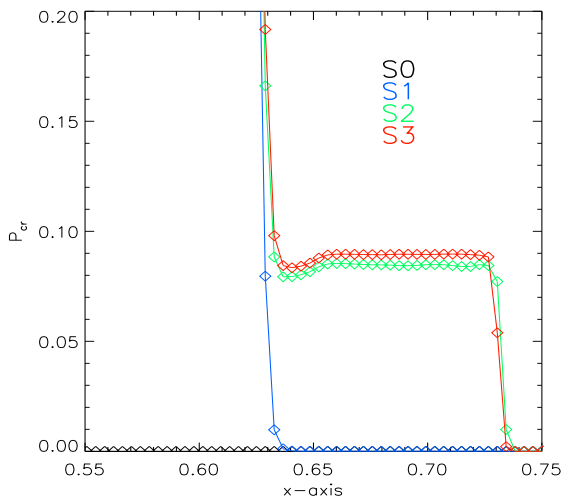}
\caption{Zoomed version of Fig.\ref{fig:tube2} for the CR pressure at the shock front. The meaning of colours
is as in Fig.\ref{fig:tube2}.}
\label{fig:tube3}
\end{figure}

\subsubsection{Dynamical feedback of CR}
\label{subsubsec:cr_feedback}

The dynamical coupling between CRs and the thermal gas is done by updating the gas pressure with the total pressure from the mixture of gas and CRs within every cell.
At each time step, the total thermal gas pressure, 
$P_{\rm g}=\rho (\gamma-1) e_{\rm g}$ (with $\gamma=5/3$) is replaced with the effective pressure in the cell, $P_{\rm eff} = P_{\rm g} + P_{\rm cr}$, where $P_{\rm cr}$ is given
in Eq.\ref{eq:pcr}. 

The effective pressure $P_{\rm eff}$ is then fed to the Riemann solver in the next time step, and the new "left" and "right" states around each cells are reconstructed
based on the new effective pressure. The momentum
equation from these states is then solved in the usual way in {\it {\small ENZO}}, and
the resulting velocity field is used to update the 
evolution of all gas quantities in the simulation, as well as of CRs energy field, $E_{\rm cr}$ (Eq.\ref{eq:CR_advect}).

In order to achieve numerical solutions without spurious
oscillations, it is crucial that all fluxes
associated with the 
total pressure, $P_{\rm eff}$, and with the CR pressure, $P_{\rm cr}$,
fed into the Riemann solver are passed through the same
slope limiters 
in order to keep the sharpness of solutions around shocks and in strong rarefaction regions (e.g. Colella \& Woodward 1984; Bryan et al. 1995). Our
tests suggests that this is particularly important when the 
energy density of CRs is of the order of $e_{\rm cr} \sim 0.1 e_{\rm g}$, or 
larger. 

Furthermore, when the "left" and "right" edge states
of the Riemann problem are computed for the solution of the fluxes,
the appropriate wave velocity is used; this implies replacing
$c_{\rm s}=\sqrt{\gamma P_{\rm g}/\rho}$ with 

\begin{equation}
c_{\rm s}'=\sqrt{\frac{\gamma P_{\rm g}+\gamma_{\rm cr} P_{\rm cr}}{\rho}}
\label{eq:cs}
\end{equation}

 in the associated routines of the  PPM algorithm, as suggested in Miniati (2007).

\subsubsection{Time stepping}
\label{subsubsec:time_step}

To ensure numerical stability in the presence of feedback from CRs, we implement 2 additional
constraints on the time stepping criteria of {\it {\small ENZO}} \citep[see][for a review]{co11}. 

First, we modify the Courant condition on the hydro-dynamical
time step by replacing the standard speed of sound with $c_{\rm s}'$ (as defined in Eq.\ref{eq:cs}):

\begin{equation}
\Delta t= {\rm min} (\frac{dx_{\rm l}}{c_{s}' \pm v}),
\end{equation} 

\noindent where $v$ is the 1--dimensional gas velocity along one axis, and the minimum is taken over the stencil of cells at each refinement level.
We need also to ensure that the internal energy correction (see Sec.\ref{subsubsec:cr_advect}), due to 
the reduced thermalization process, is not larger than a fraction of 
the gas energy within the cell, since this would cause numerical instabilities.
We achieve this by imposing:

\begin{equation}
\Delta t_{\rm cr} \leq \epsilon_{\rm cr} \cdot e_{\rm g} \cdot \frac{\Delta_{\rm l}}{\phi_{\rm cr}},
\end{equation}

\noindent which follows from the requirement that $\epsilon_{\rm cr}  e_{\rm g}\geq e_{\rm cr}$ at each cell. 
Our tests suggest that $\epsilon_{\rm cr}=0.02-0.04$ is a suitable choice to ensure that the modification
of the thermal energy in the post-shock is not dramatic during each time step, also in the case of the complex 3--D flows (which may cause 
fluxes of accelerated CRs across several faces of a cell).  
Compared to the standard cosmological simulations at fixed grid resolution, the use of these additional
time stepping criteria usually produce an increase of $\sim 15$ percent in the total CPU time by the
end of runs.

\bigskip

\section{Tests in one dimension.}
\label{sec:test1}

\subsection{Shock Tube with pre-existing Cosmic Rays}
\label{subsec:tube_first}

We performed one-dimensional shock tube tests assuming a mixture of CRs and thermal gas in the initial setup, and investigating the behaviour
of our scheme at different Mach numbers.

The parameters of the first shock tube test are 
identical to those reported in Miniati (2007):
the initial left state is filled with a mixture
of gas and CRs, with pressure (in code units) $P_{\rm g,l}=10$, $P_{\rm cr,l}=6$,
and density $\rho_{\rm l}=9$, while all the fields
on the right state are set to 1 ($\gamma_{\rm cr}=4/3$ is
assumed everywhere for CRs).

In Fig.\ref{fig:tube1} we show the comparison of our test
with the numerical solution of Miniati (2007), at
approximately the same epoch. In  this case, all physical modules in our version of {\it {\small ENZO}} with
Cosmic Rays are activated: advection, injection, pressure feedback and reduced thermalization at the post-shock.
The agreement of our run with the results of Miniati (2007) is excellent, with a constant total pressure level in the post-shock region. 
The contact discontinuity at the rarefaction wave is reconstructed in a few cells (which is common
in the PPM method), while the transition of the shock is reconstructed in $\sim 2$ cells.

The numerical implementation of our CR code also allows us to investigate the role of each
physical module separately. 
We show the effect of each module in Fig.\ref{fig:tube2}, where the same initial conditions
of the previous shock tube were re-simulated by modelling: only the advection of CRs is modelled (run  {\it S0}); the advection of CRs and their pressure feedback
onto the gas component ({\it S1}); the injection of new CR energy at the shock 
({\it S2}); the whole set of CR modules ({\it S3}). 

When we allow CRs to have pressure feedback on the gas ({\it S1}), the total pressure
of the system  is increased and the behaviour of gas density, velocity and of the total
pressure are scaled-up versions of those in {\it S0}. However, the behaviour of
gas pressure at the shock front is modified by the presence of the CR pressure. 
The injection of CR energy at the shock front in {\it S2} (here $\eta \approx 0.03$) slightly increases the total pressure at the shock compared to {\it S1}. 
When including the reduced thermalization ({\it S3}), the total pressure at the shock front is smaller compared to {\it S2}, and as a result the 
speed of the shock front is decreased. 
Figure \ref{fig:tube3} shows a close-up of the behaviour of CR pressure at the shock front. Only very small oscillations (at the percent level) are present in the post-shock
solution of our CR fluid.

In all runs where the CRs are allowed 
to exert a pressure on the gas ({\it S1}, {\it S2} and {\it S3}), the
solution in the rarefaction region does not vary. This
confirms that the treatment of CR acceleration at the shock
does not affect any other region in the shock tube. 

\begin{figure}
\includegraphics[width=0.45\textwidth]{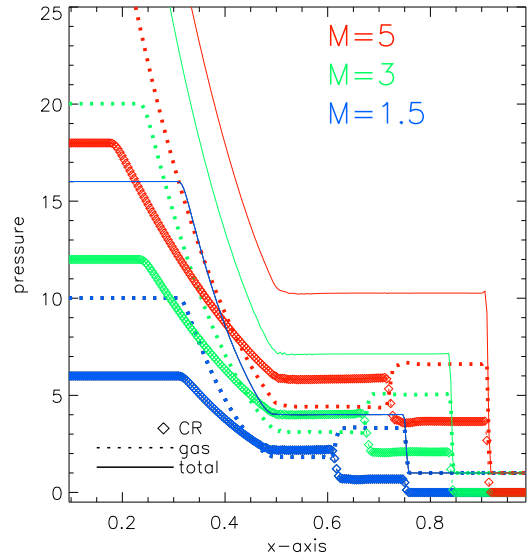}
\caption{Behaviour of the gas pressure (dots), CR pressure (diamonds) and total pressure (solid lines) for shock
tube tests leading to a $M \approx 1.5$ (blue), $M\approx 3$ (green) and $M \approx 5$ (red) shocks. Only the
values of CR pressure are shown for clarity; the x-axis is in units of the total box length.} 
\label{fig:tube4}
\end{figure}

In a second set of shock tube tests, 
we varied the initial conditions by increasing the gas pressure on the
left side in order to produce stronger shocks. 
In Fig.\ref{fig:tube4} we show the results for shock tubes
leading to $M \approx 1.5$, $3$ and $5$ and assuming $\eta \approx 0.3$ to highlight the 
effect of CR injection at the shock. Even for the strongest shock, the post-shock total 
pressure shows no significant spurious oscillation, and the shock edge is still reconstructed in
2-3 cells at most.

\begin{figure}
\includegraphics[width=0.45\textwidth]{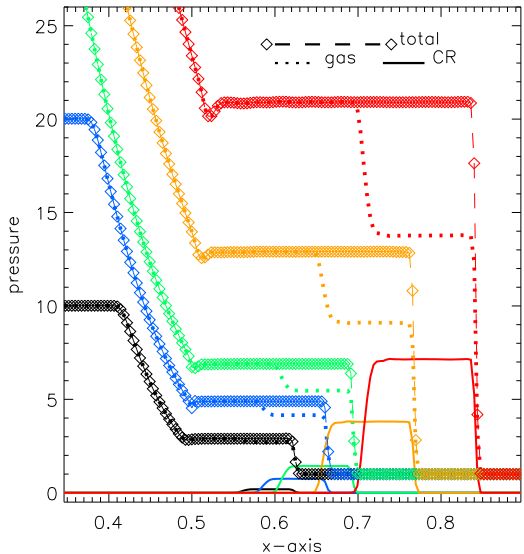}
\caption{Behaviour of the gas pressure (dots), CR pressure (diamonds) and total pressure (solid lines) for several shock
tube tests without pre-existing CR energy. Only the cell-wise
values for the total pressure are shown for clarity.} 
\label{fig:tube5}
\end{figure}

\subsection{Shock Tube without pre-existing Cosmic Rays}
\label{subsec:tube_nocr}

Next, we focus on the injection of CR energy at the shock front by simulating
similar 1-D shock tubes as in the previous case, but assuming no pre-existing CR energy at the beginning of the simulation.
Also, we increased the values of initial
gas pressure on the left side of the box ($P_{\rm g,l}=$10, 20, 30, 60 and 120).
In these tests, we let $\eta(M)$ vary with the shock Mach number according to the
model of Kang \& Jones (2007), which we will use in our cosmological runs.

The results are shown in Fig.\ref{fig:tube5}.
At the strongest shocks, the freshly injected CR pressure becomes comparable to the
post-shock gas pressure ($P_{\rm cr} \sim 0.3 P_{\rm g}$). The total pressure
behind the shock still remains constant, despite a small amount of oscillations
of numerical nature. Again, the shock edge is reconstructed within 2-3 cells at most in all
tests.

To avoid any smoothing of the shock region, which may
cause spurious effects when the non-linearity of 
CR acceleration schemes is coupled to the system, Miniati (2007) proposed the 
application of a hybrid Godunov-Glimm's scheme to ensure the sharpest reconstruction
of shocks in the simulation.

In order to limit the effect of a broadened shock profile on the estimate of the injected CR
energy in all our tests here and in the production runs (Sec.\ref{sec:results}), our run-time shock finder (Sec.\ref{subsec:shock}) measures Mach numbers in the
simulations across 3 cells. This was motivated by our previous work on
cosmological shock waves in {\it {\small ENZO}} simulations (Vazza, Brunetti \& Gheller 2009), where we studied
the best stencil of cells needed to correctly capture shock waves in the PPM version of {\it {\small ENZO}}.
With this approach, the effect of numerical smearing of reconstructed shocks only  has a very
limited impact on the estimated injection efficiency of CRs, since $M$ is always measured with a stencil
of cells close to the (small) numerical smearing of shocks in the PPM method.

\begin{figure*}
\includegraphics[width=0.95\textwidth,height=0.76\textwidth]{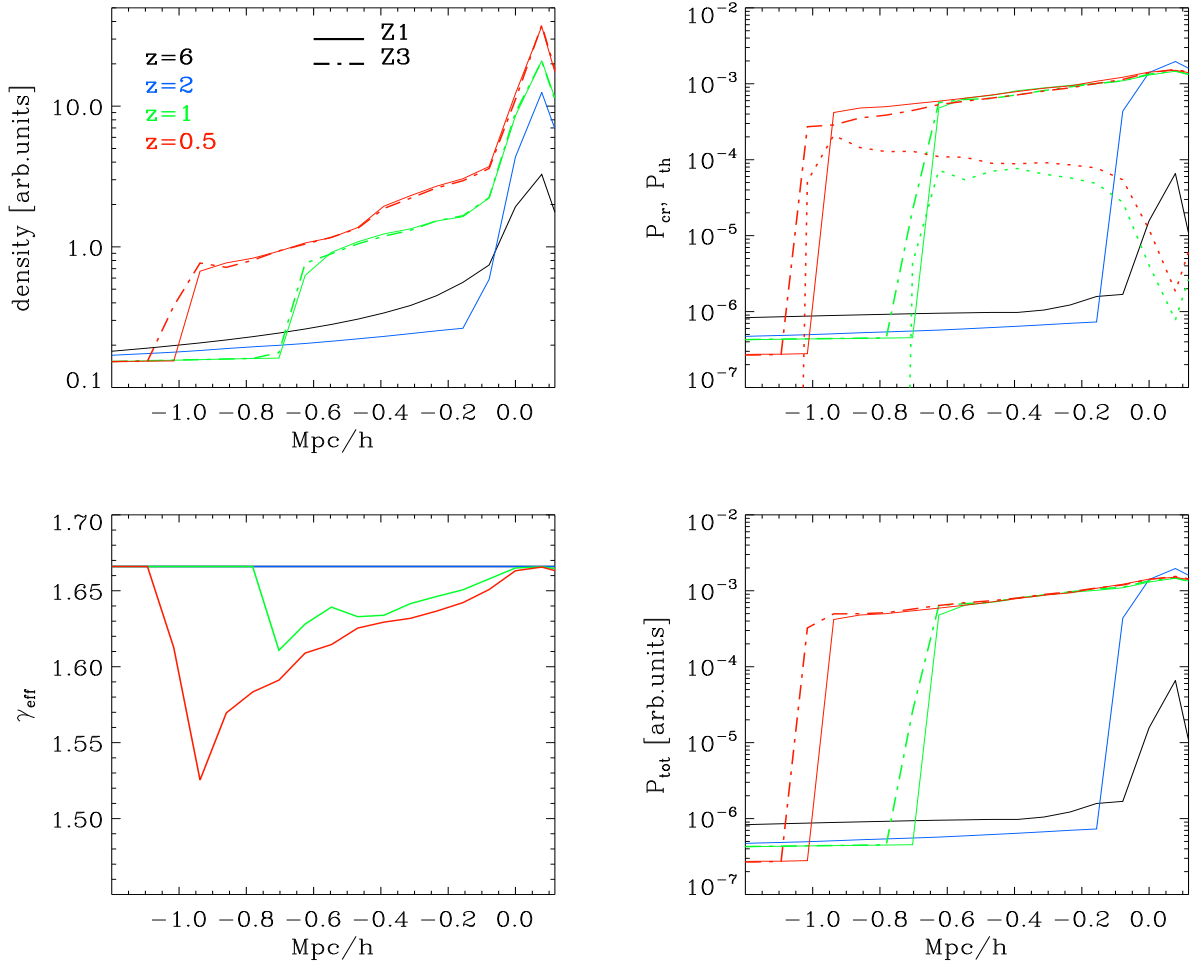}
\caption{Time evolution of the profiles of gas density, 
gas and CR pressure (dotted lines in the second panel), effective adiabatic index (bottom left) and total pressure (bottom right)  for the Zeldovich collapse test run with no CR physics (solid lines, run {\it Z1}) and with CR physics (dot-dash, run {\it Z3}).}
\label{fig:zeld1}
\end{figure*}

\begin{figure}
\includegraphics[width=0.45\textwidth]{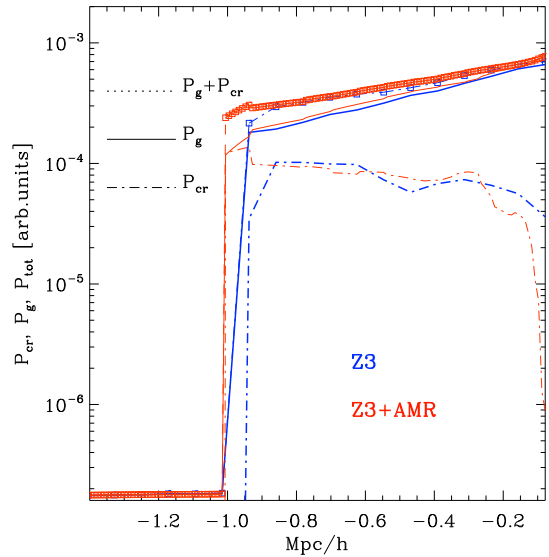}
\caption{Comparison of the total pressure of the Zeldovich collapse test in {\it Z3} employing fixed resolution and of the
{\it Z3} with 4 levels of AMR. The different lines show: gas pressure (solid), CR pressure (dot-dashed) and the total pressure (dotted).}
\label{fig:zeld3}
\end{figure}

\subsection{Zeldovich Pancake}
\label{subsec:zeld}

In order to test our implementation of cosmic ray physics in a simple cosmological framework, we run
a set of Zeldovich Pancake test and compare the results with and without the dynamical
role of CR energy injected at shocks.
The initial
conditions ($z_{\rm i}=50$ in our case) for this test assume a purely baryonic Universe, with a uniform initial
pressure and density, and a converging velocity field with $v=0$ at the centre of the domain.
With this setup we simulate the evolution of a co-moving box of $L_{\rm box}=20$ Mpc/h ($h=0.72$) with a grid of 256 cells. No CR energy is present in the box at the start of the 
simulation. 
Here
results are presented for the simple collapse test (run {\it Z1}), and for the 
full set of physical modules for CRs ({\it Z3}). 
In all cases the acceleration efficiency is the one of \citet{kj07}. Additionally,  in order
to test the reliability of our method with adaptive mesh refinement, an
AMR run with a 16 times higher spatial resolution  has been performed (run {\it Z3+AMR}).

The panels in Figure \ref{fig:zeld1} show the evolution of gas
density, gas pressure, CR pressure, total pressure and total energy
in runs {\it Z1} and {\it Z3}. 
By the end of the runs the profile of the "pancake" is very similar in 
all cases, with sizable differences at the percent level which
are evident for $z<2$, after enough CR energy has been injected by
the expanding shocks.
The gas density and gas pressure of {\it Z3} are slightly smaller inside the structure and 
slightly larger in the external regions (at $\sim -0.8-1.2$ Mpc/h), while
the total pressure profile is very similar in the two cases, with the
exception of the outer region where run {\it Z3} shows an excess. 
The reason for this behaviour is that, when output at equal times is compared in runs {\it Z1} and {\it Z3},
the outer shocks in the simulation with CRs are found to have expanded to a slightly
larger distance, due to their enhanced total pressure.  The combination of 
the softer equation of state of the gas+CR composite adopted in {\it Z3} at the outer regions, and the total larger pressure jumps felt by the gas accreted onto the pancake makes
this mechanism very stable, since both effects concur to 
have a more efficient injection of CRs in the system, due to the increase of the Mach 
number of external shocks. We will discuss this mechanism again in Sec.4 and in the Appendix, with the
results of fully  cosmological  runs in 3 dimensions.

By the end of this run, the pressure ratio between CRs and gas 
is  $P_{\rm cr}/P_{\rm g} \sim 0.6$
at the external regions and $P_{\rm cr}/P_{\rm g} \sim 0.05-0.1$ in the
innermost region of the pancake.

In Fig.\ref{fig:zeld3} we show the effect of AMR (4 levels of refinement, triggered by the local
gas overdensity) on the same collapse test. 
The solution recovered by the end of the run is quite similar to {\it Z3}, provided     that 
the larger spatial resolution here enables a sharper modelling of the
outer shock, and leads to a larger injection of CRs. These two effects 
modify the shock speed in a non-linear way, making the synchronization of the 
two runs not trivial. 
The energy ratio measured at the outer region is larger
compared to {\it Z3}, yielding $P_{\rm cr}/P_{\rm g} \sim 0.5$. 
Based on the profile of $P_{\rm g}$ and $P_{\rm cr}$ along
the pancake, we conclude that the previous results at coarser resolution are
stable against such a dramatic increase of resolution, even if the detailed
distribution of $P_{\rm cr}$ at the shock edge may slightly vary with resolution. 
As in the previous case, this difference can be partially ascribed to small timing issues in 
the comparison of the two runs, and partially to the slightly enhanced amount of CRs injection
measured at higher resolution.

\begin{table}
\label{tab:tab1}
\caption{Numerical and cosmological parameters for the runs employing fixed mesh resolution. The columns lists:
ID of the run; mass resolution
for DM;  gas spatial resolution; number of grid cells. The last column reports mnemonics explaining the physical 
setups used in the run. "{\it A}"=advection of CR; "{\it I} "=injection of CR at shocks; "{\it T}"=reduced 
thermalization at the post-shock; "{\it P}"=pressure feedback of CRs. "$z_{\rm cr}=1,29$" mean that the injection
in the runs was started at $z=1$ and $z=29$, respectively (or at $z=6$  elsewhere). "{\it no-re.}" means that
the run does not model a re-ionization background (otherwise considered). "$\eta=0.01$" and "$\eta=0.01$" mean
that the acceleration efficiency in the runs was kept fixed at these two values, independent on the measured Mach
number (the $\eta(M)$ efficiency of Kang \& Jones (2007) was used elsewhere).  
"{\it $M_{\rm cr}$}"  means that the Mach number used to compute $\eta(M)$ is the one from the total gas+CRs pressure
jump, instead of the gas pressure jump only. "$\rho_{\rm cr}=10^{-3},10$" means that we assumed minimum overdensity
of $10^{-3}$ or $10$ the background critical gas density to inject CRs ($\rho_{\rm cr}=0.1$ elsewhere).}
\begin{tabular}{c|c|c|c|c}
ID & $M_{\rm dm}$ & $\Delta_{\rm x} $  & $N_{\rm grid}$ & mnemonic \\ 
 & $[M_{\odot}/h]$ & [kpc/h] & [cells] &  \\ 
\hline
Run1 & $ 5.1 \cdot 10^{9}$ & 448 &  $128^{3}$ & AI\\
Run2 & $ 5.1 \cdot 10^{9}$ & 448 & $128^{3}$ &  AIP\\
Run3 & $ 5.1 \cdot 10^{9}$ & 448 &  $128^{3}$ & AIPT \\
Run3\_l  & $ 4.0 \cdot 10^{10}$ & 996 & $64^{3}$ & AIPT \\
Run3\_m  & $ 6.4 \cdot 10^{8}$ & 224 & $256^{3}$ & AIPT \\
Run1\_h & $ 8 \cdot 10^{7}$ & 112 &  $512^{3}$ & AI \\
Run3\_h & $ 8 \cdot 10^{7}$ & 112 &  $512^{3}$ & AIPT\\
Run3\_nore & $ 5.1 \cdot 10^{9}$ & 448 &  $128^{3}$ & AIPT, no-re. \\
Run3\_z1 & $ 5.1 \cdot 10^{9}$ & 448 &  $128^{3}$ & AIPT, $z_{\rm cr}=1$ \\
Run3\_z29 & $ 5.1 \cdot 10^{9}$ & 448 & $128^{3}$ & AIPT, $z_{\rm cr}=29$ \\
Run3\_eta001  & $ 5.1 \cdot 10^{9}$ & 448 & $128^{3}$ & AIPT, $\eta=0.01$.\\
Run3\_eta01  & $ 5.1 \cdot 10^{9}$ & 448 & $128^{3}$ & AIPT, $\eta=0.1$.\\
Run3\_d0001 & $ 5.1 \cdot 10^{9}$ & 448 & $128^{3}$ & AIPT, $\rho_{\rm cr}=10^{-3}$\\
Run3\_d10 & $ 5.1 \cdot 10^{9}$ & 448 & $128^{3}$ & AIPT, $\rho_{\rm cr}=10$\\
Run3\_mcr & $ 5.1 \cdot 10^{9}$ & 448 & $128^{3}$ & AIPT, $M_{\rm cr}$, $\eta=0.1$\\

 \end{tabular}
\end{table}

\begin{table}
\label{tab:clusters}
\caption{Main characteristics of the simulated clusters. 
Column 1: identification number. Columns 2:
total mass ($M_{\rm TOT}= M_{\rm DM}+M_{\rm gas}$) inside the virial radius.  Columns 3: virial radius, $R_{\rm v} \approx 0.7 R_{\rm 200}$. Column 4: 
dynamical state at $z=0$: MM=major merger cluster; ME=merging cluster, RE=relaxed cluster (see Sect.\ref{subsec:amr} for further details).}
\begin{center}
\begin{tabular}{c|c|c|c|c}
ID & $M_{\rm TOT}[10^{14} M_{\odot}]$ & $R_{\rm v}$[kpc] & dynamical state\\
\hline

E1 & 11.20 & 2670 & MM\\
E7 & 6.52 & 2194  & ME \\
E25 & 6.55  &  2369 & MM \\ 
H1 & 3.10 & 1890 & RE    \\  
H3 & 2.95 & 1710 & ME   \\  
H5 & 2.41 & 1703 & MM    \\  
H7 & 2.14 & 1410 & RE   \\   

\end{tabular}
\label{tab:tab2}
\end{center}
\end{table}

\begin{figure*}
\includegraphics[width=0.45\textwidth,height=0.37\textwidth]{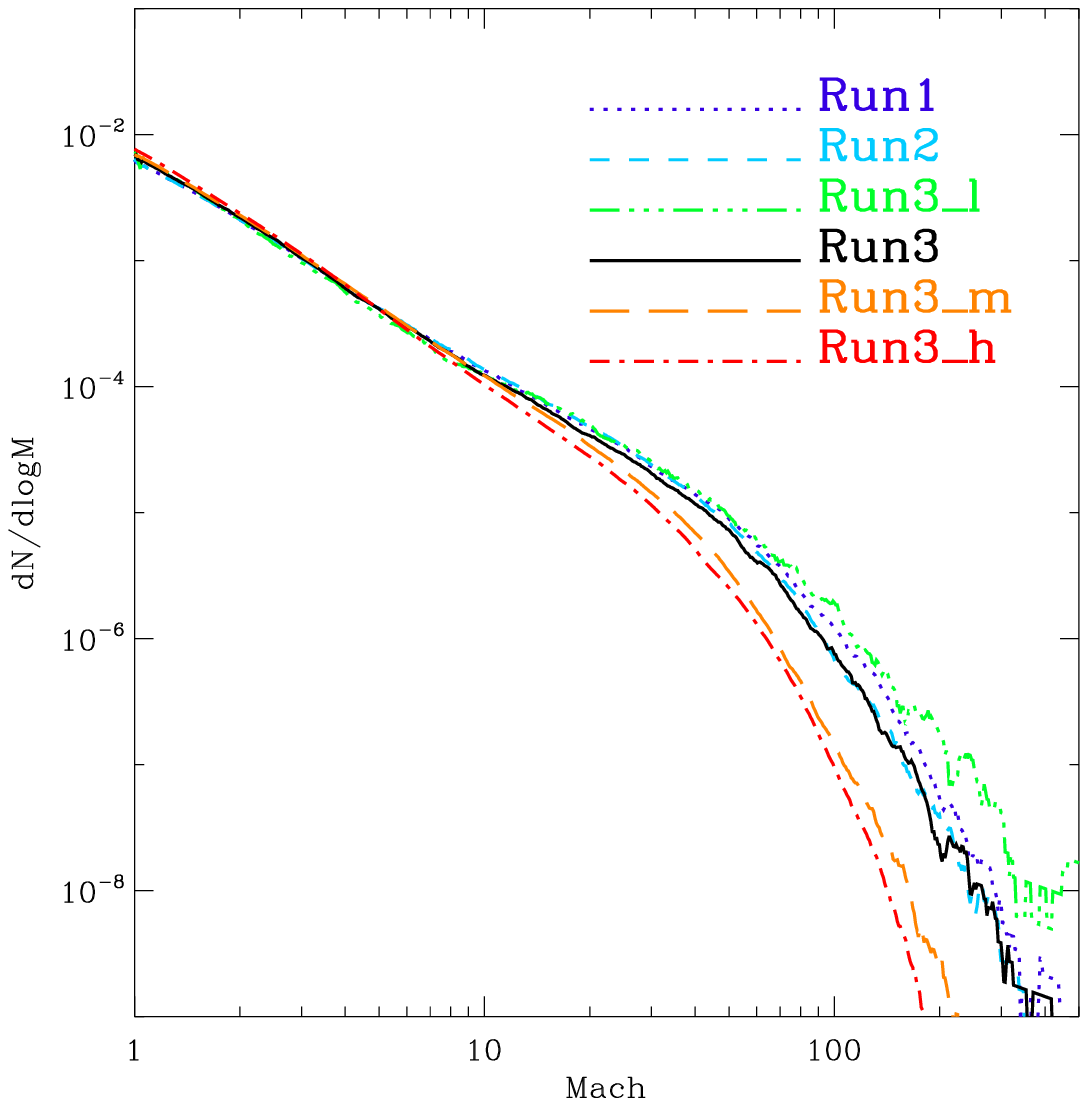}
\includegraphics[width=0.45\textwidth,height=0.37\textwidth]{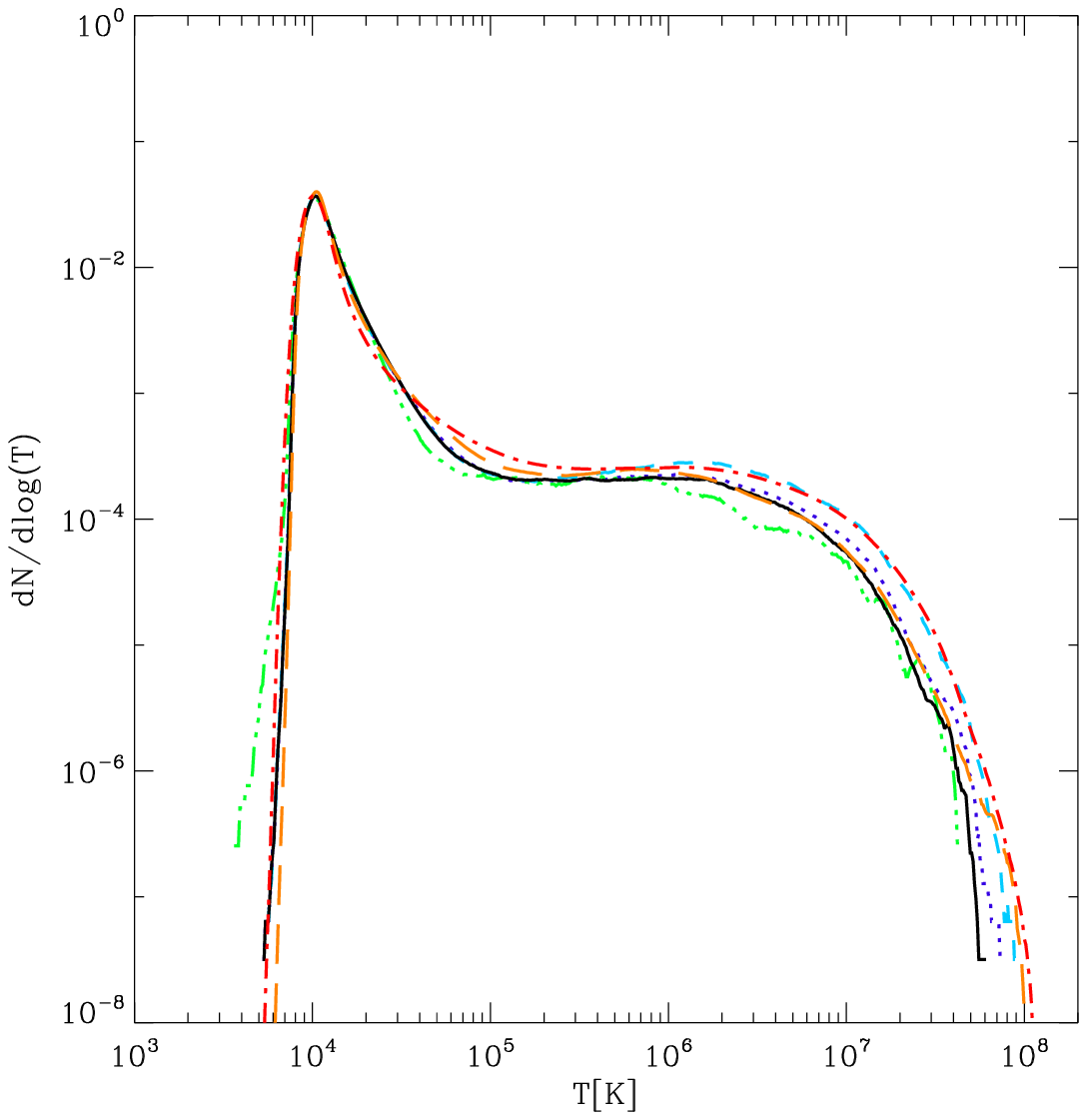}
\includegraphics[width=0.45\textwidth,height=0.37\textwidth]{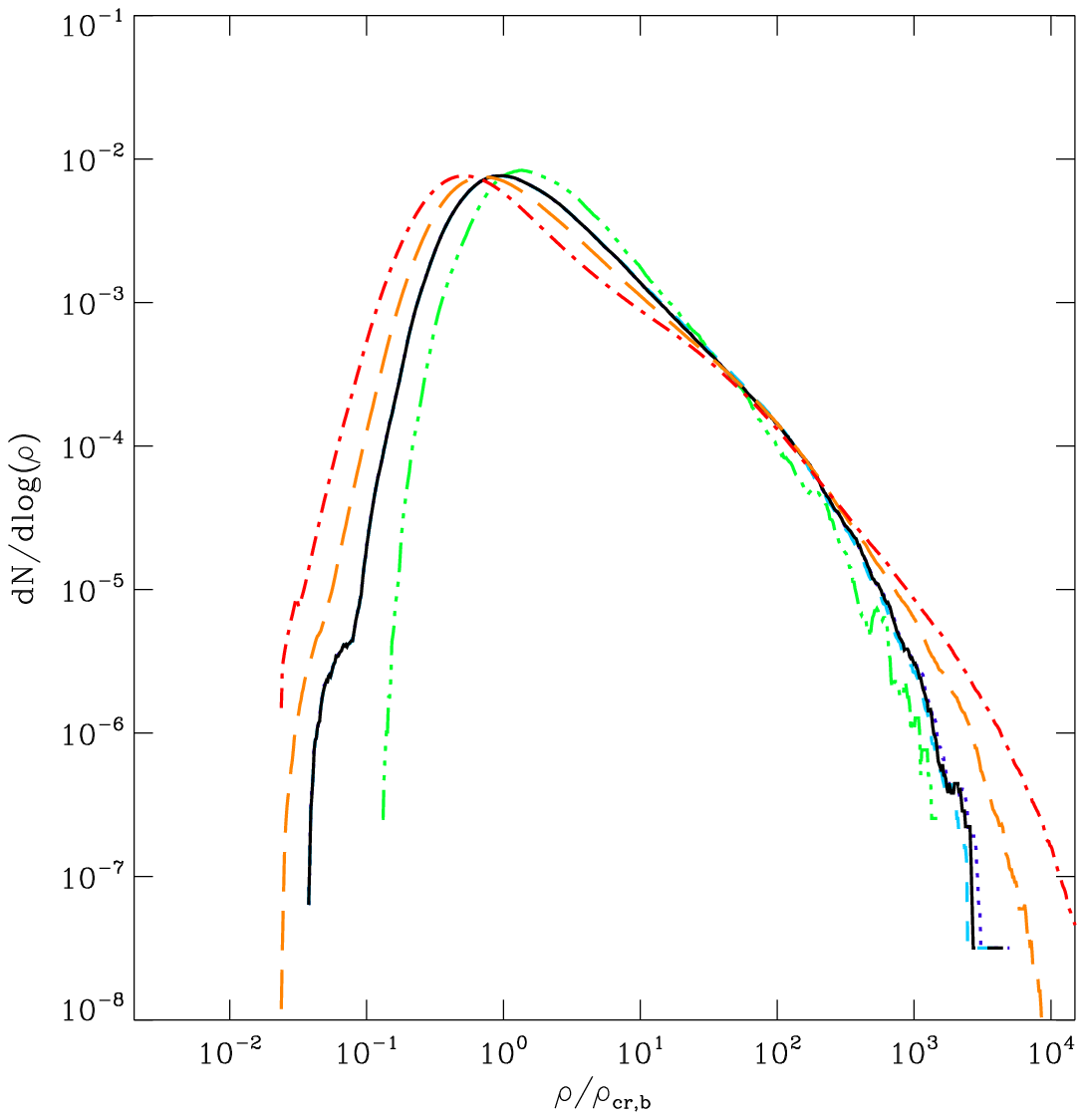}
\includegraphics[width=0.45\textwidth,height=0.37\textwidth]{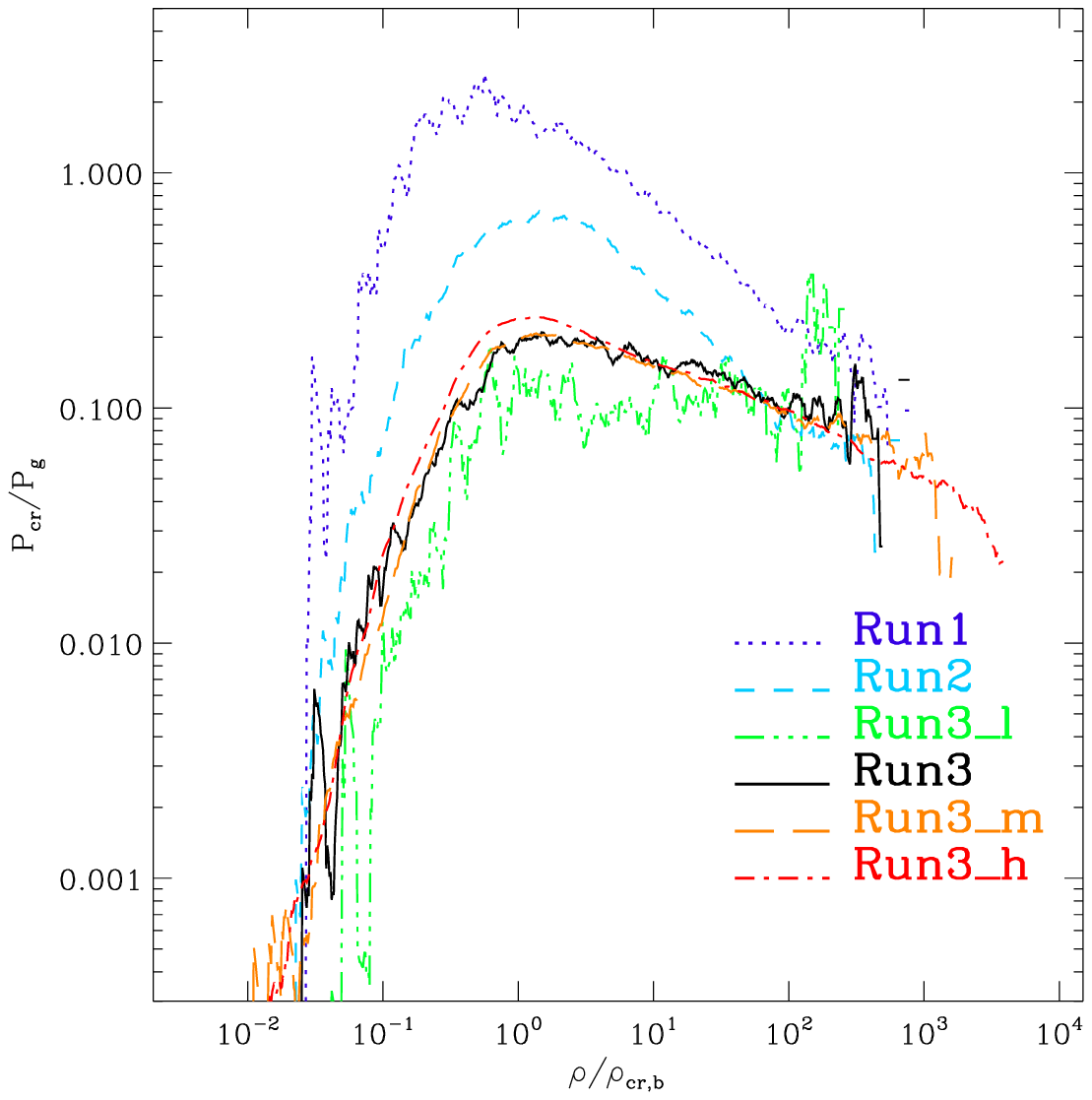}
\caption{Comparison for the distributions of Mach number (upper left panel), gas temperature (upper right), gas density (lower left) 
and $P_{\rm cr}/P_{\rm g}$ as a function of gas over-density (lower right) for runs Run1, Run2, Run3, Run3\_l, Run3\_m and Run3\_h.}
\label{fig:lls_scaling}
\end{figure*}

\begin{figure}
\includegraphics[width=0.49\textwidth]{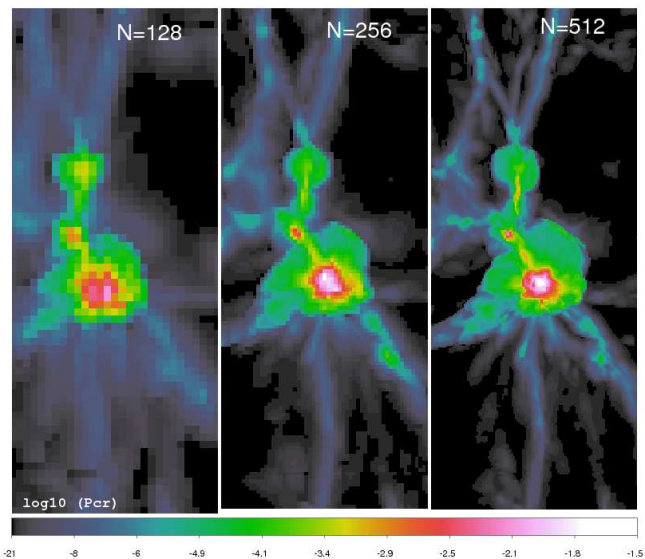}
\caption{Maps of CR pressure as a function of resolution, for a slice of $20 \times 60$ Mpc and width 448 kpc/h, centred
on the same massive galaxy cluster in the simulated volume (Run3, Run3\_m and Run3\_h are shown).}
\label{fig:lls3}
\end{figure}

\bigskip

\begin{figure}
\includegraphics[width=0.45\textwidth]{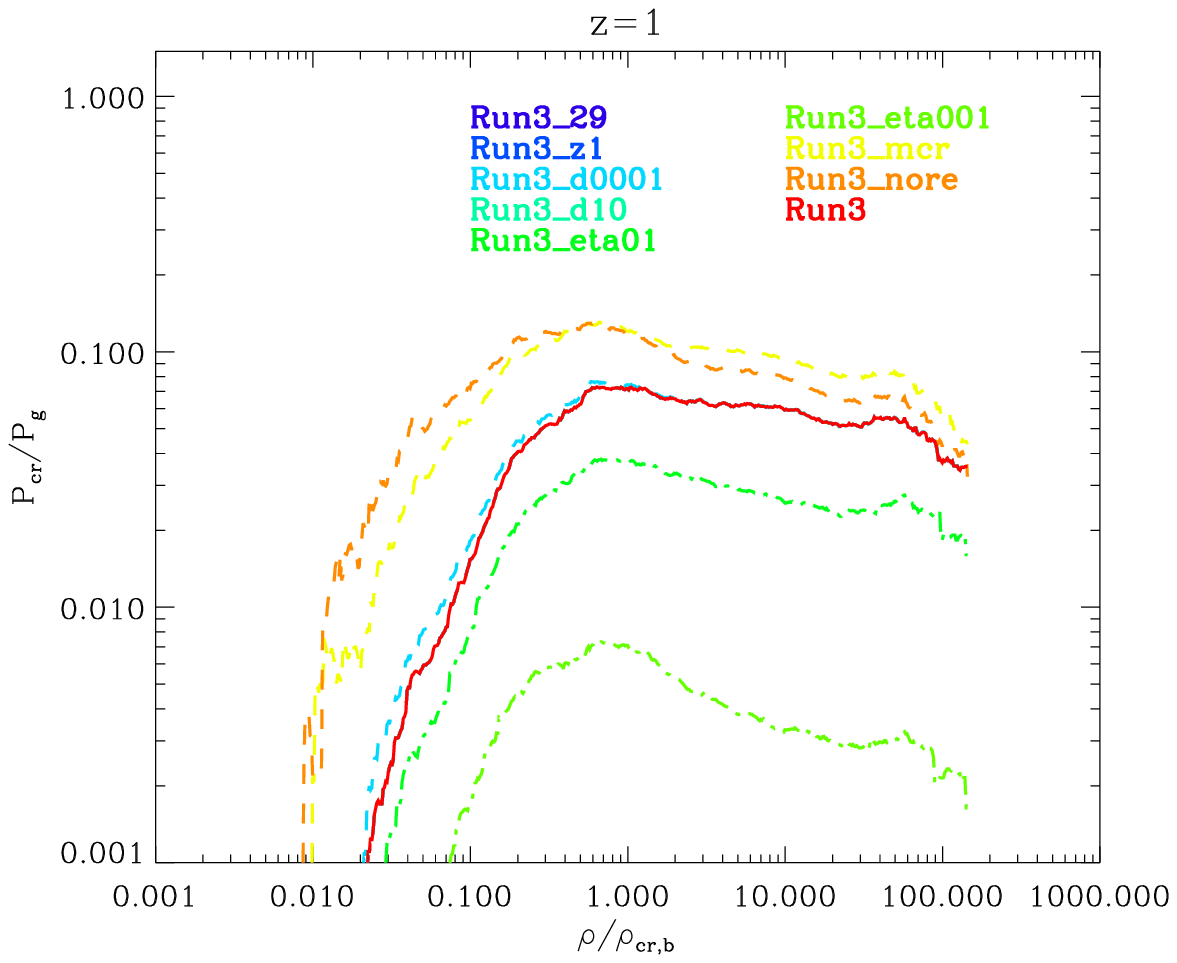}
\includegraphics[width=0.45\textwidth]{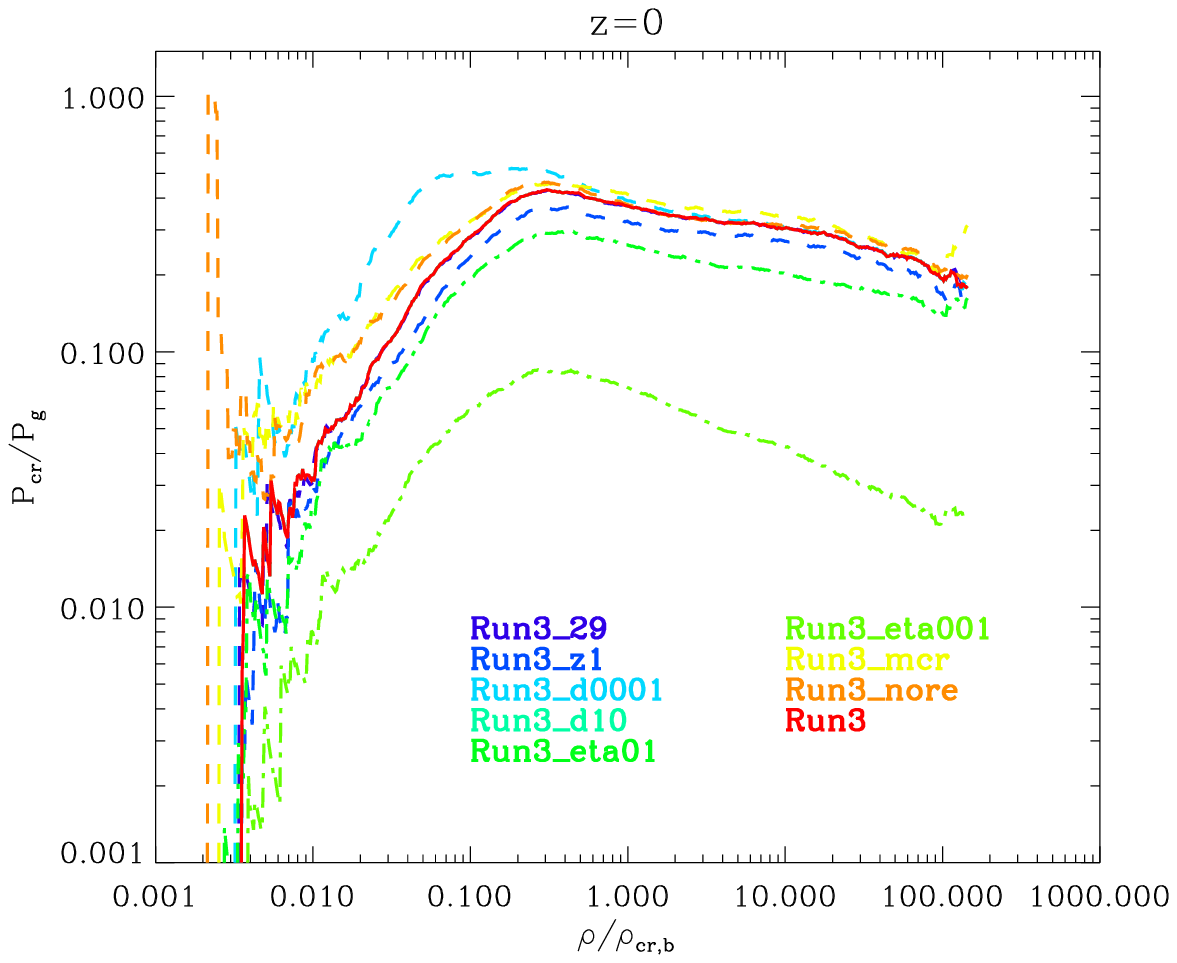}
\caption{Average pressure ratio between CRs and thermal gas as function
of the gas over-density in our runs at fixed grid resolution for
$z=1$ (top) and $z=0$ (bottom).}
\label{fig:lls_scaling_eff}
\end{figure}

\begin{figure}
\includegraphics[width=0.49\textwidth,height=0.31\textheight]{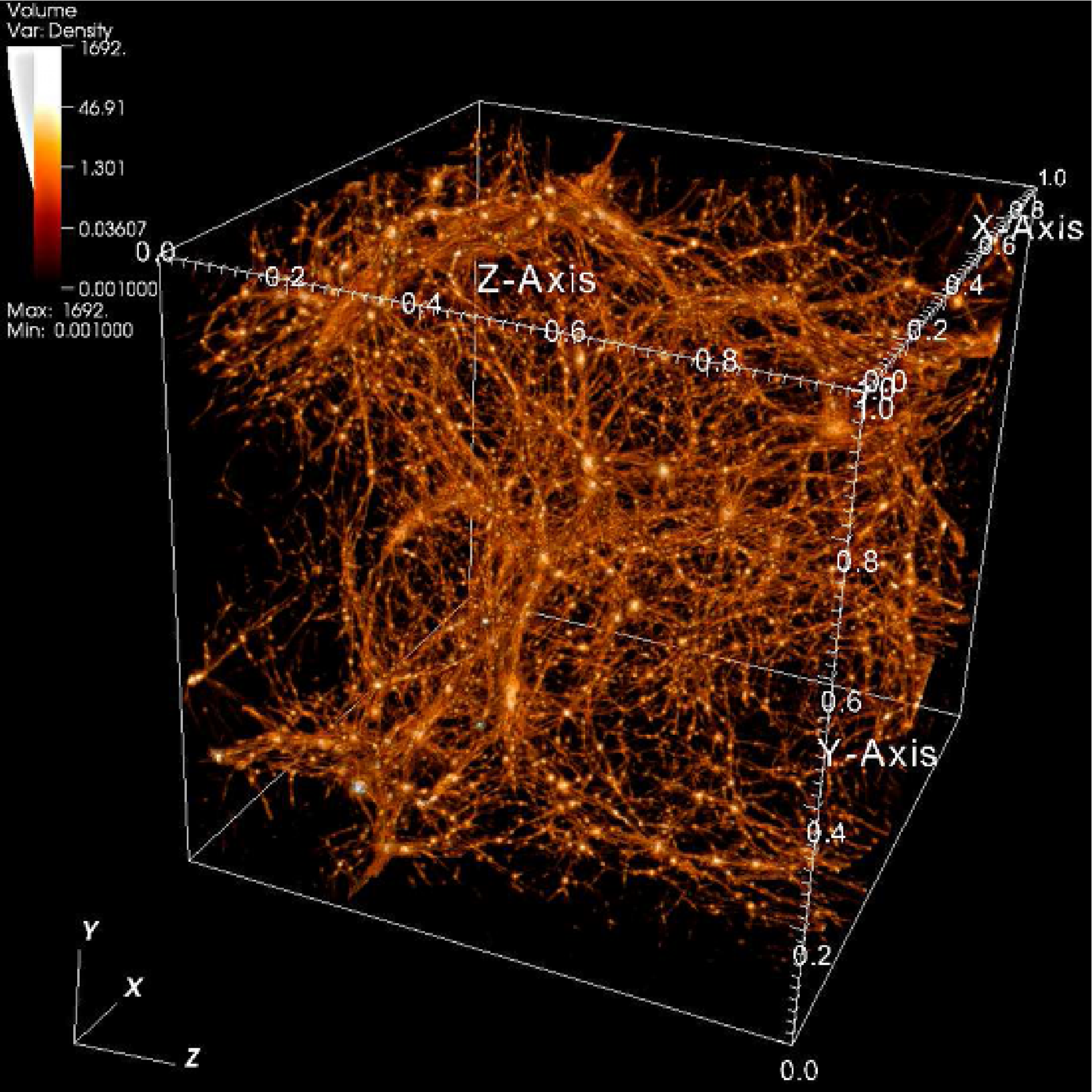}
\includegraphics[width=0.49\textwidth,height=0.31\textheight]{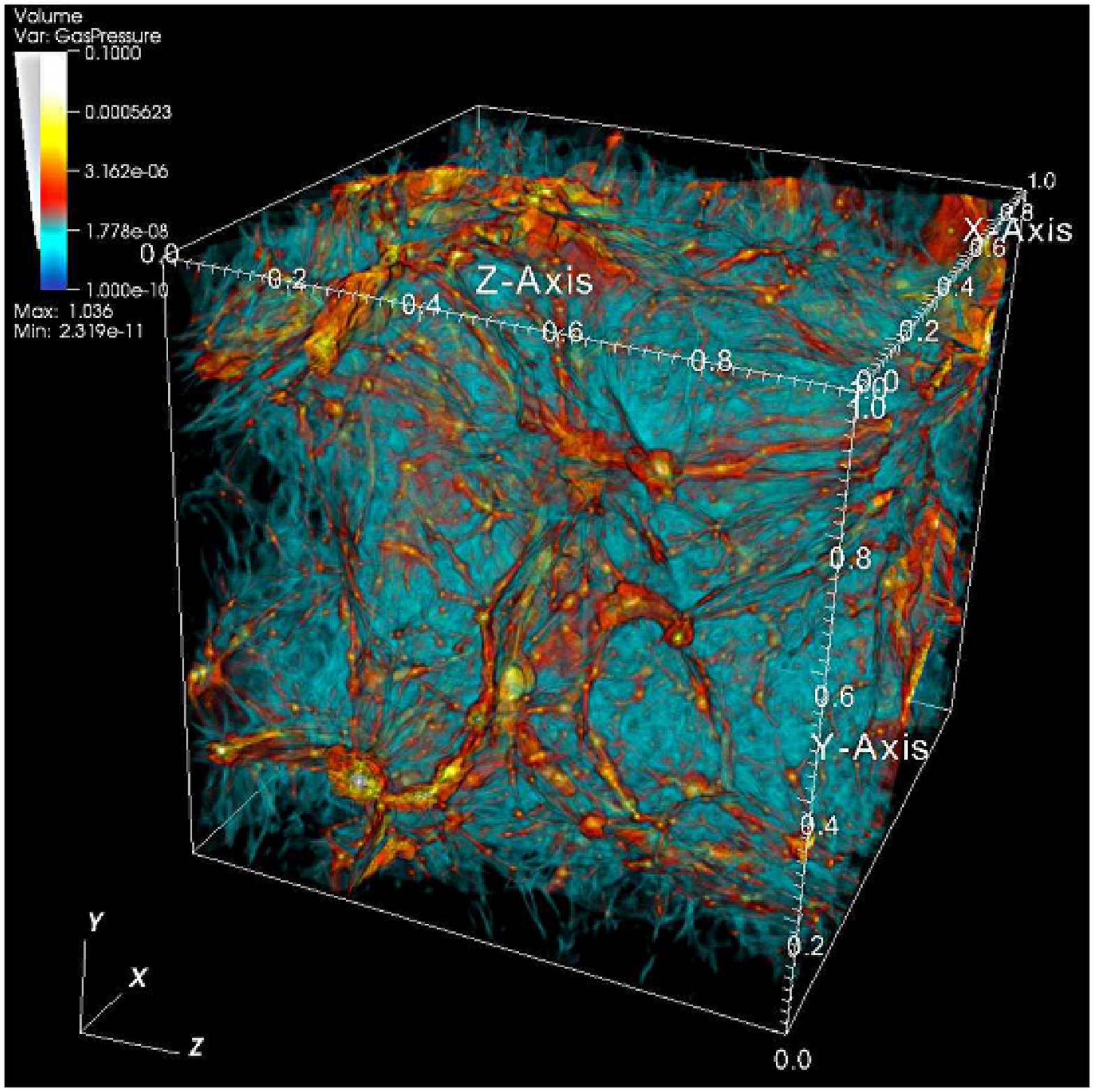}
\includegraphics[width=0.49\textwidth,height=0.31\textheight]{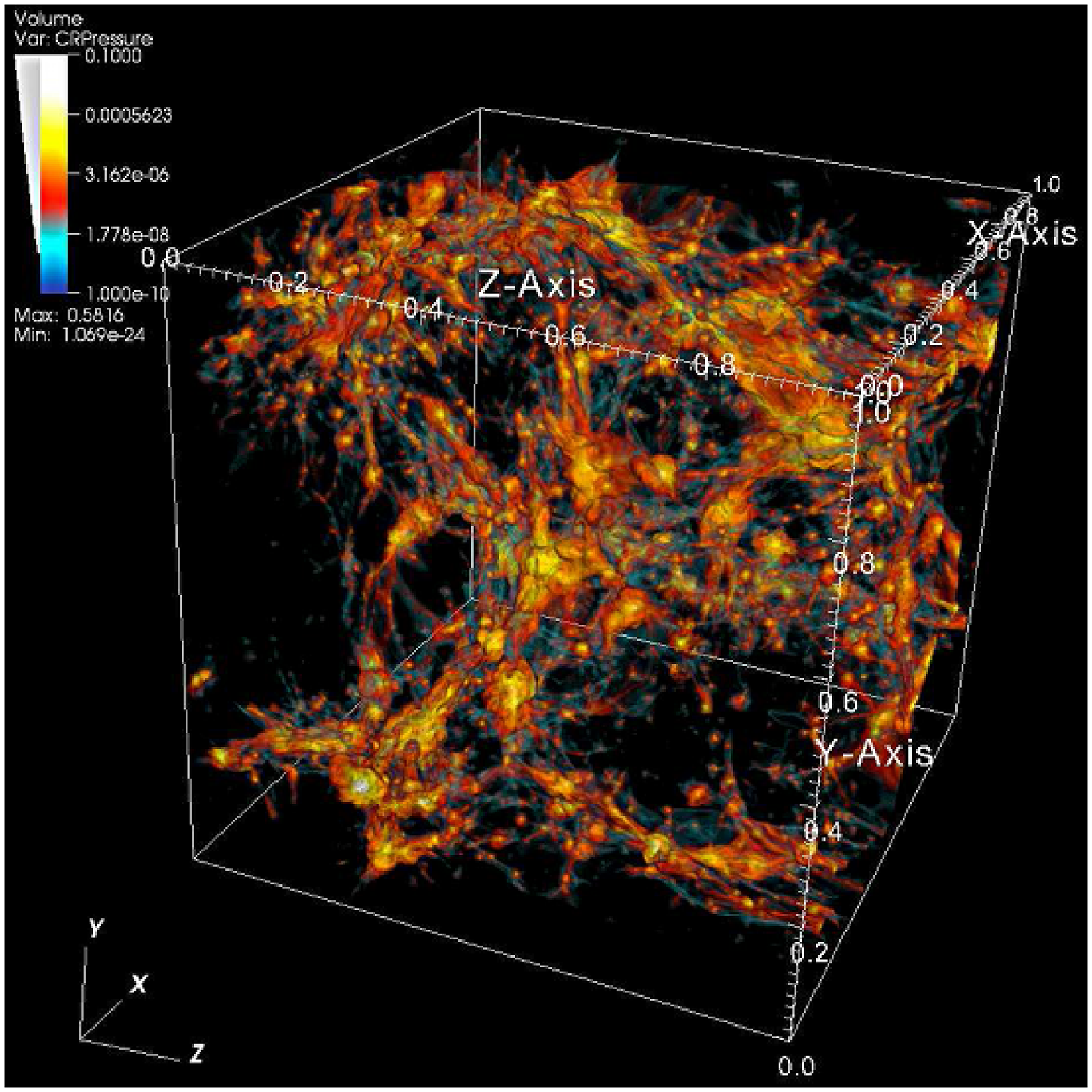}
\caption{From top to bottom: 3--D volume rendering of gas density (in 
$[\rho/\rho_{\rm cr,b}]$), 
$P_{\rm g}$  and $P_{\rm cr}$ (arbitrary code units)
 for $(80 {\rm Mpc})^{3}$ in Run3\_h at $z=0$.
Rendering performed with VISIT 2.3.1 ({\it https://wci.llnl.gov/codes/visit)}.}
\label{fig:maps_projected}
\end{figure}

\begin{figure}
\includegraphics[width=0.4\textwidth,height=0.6\textwidth]{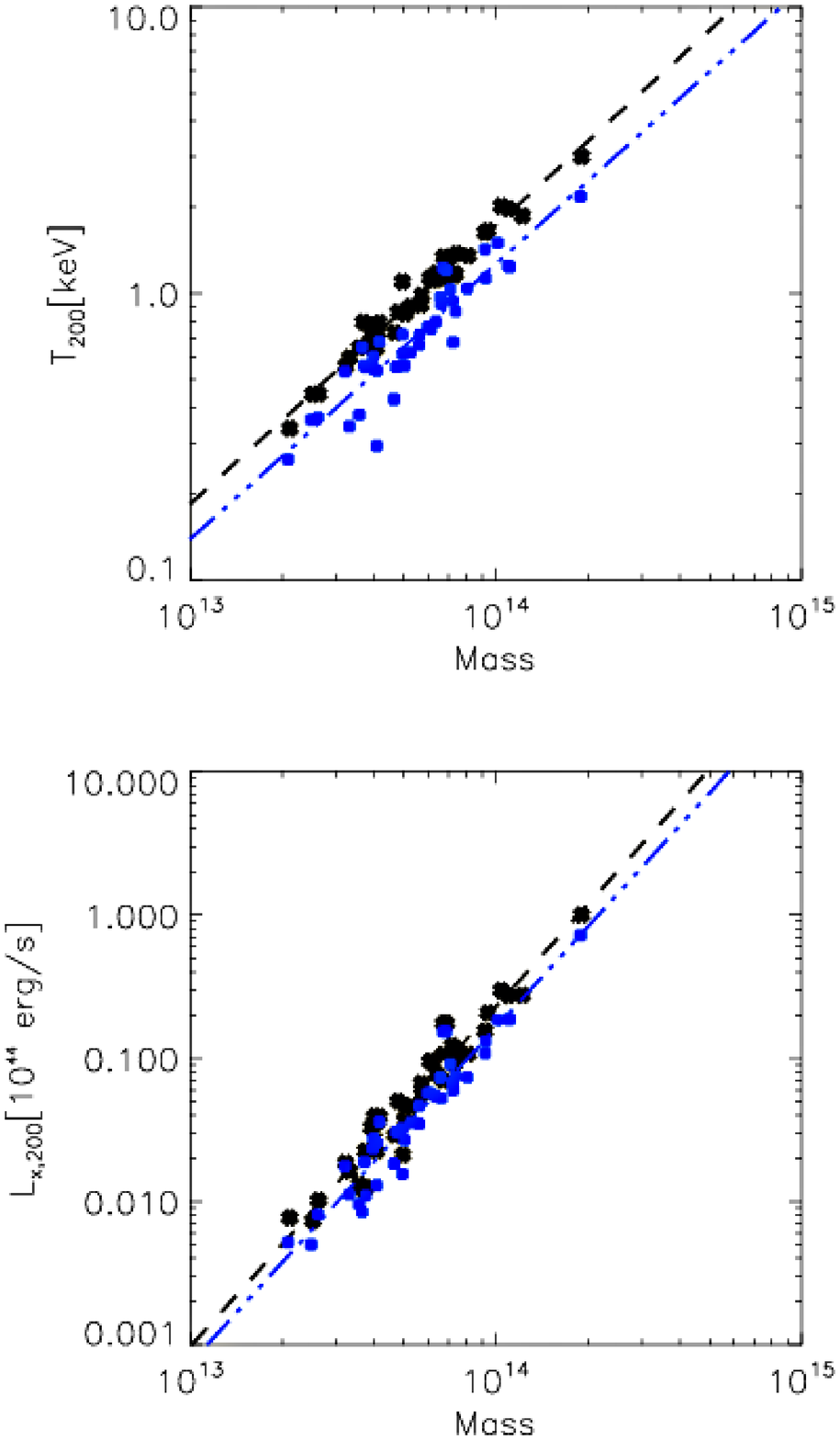}
\includegraphics[width=0.4\textwidth,height=0.3\textwidth]{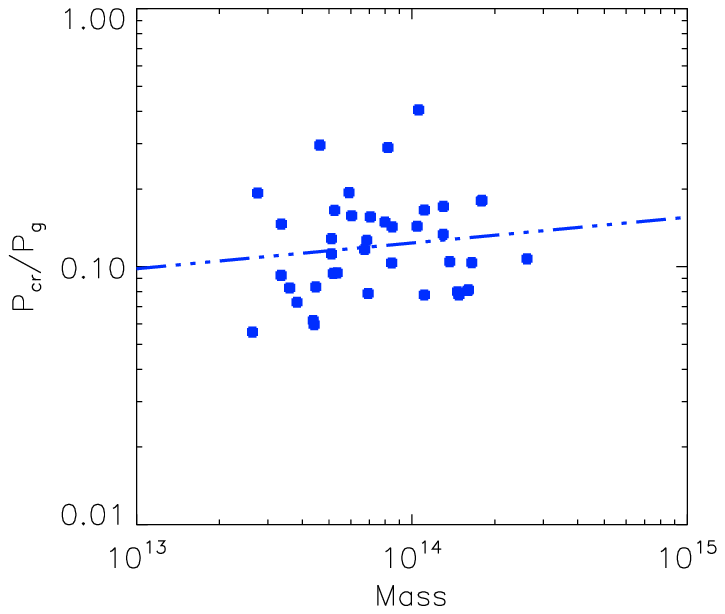}
\caption{Scaling laws for the 40 most massive halos in Run1\_h (black dots) and Run3\_h (blue). 
Top panel: scaling between the total cluster mass, $M_{\rm 200}$ (in $M_{\odot}$ and the mass-weighted temperature,
 $T_{\rm 200}$. Central panel: scaling between $M_{\rm 200}$
and the total X-ray bolometric luminosity inside $R_{\rm 200}$, $L_{\rm X,200}$. The straight
 lines show the best-fit for each sample. Bottom panel: average pressure ratio between CRs 
 and gas for all clusters in Run3\_h.}
\label{fig:scaling}
\end{figure}

\section{Results in cosmological simulations}
\label{sec:results}

We produced a large set of cosmological simulations in order to study the evolution of 
CRs accelerated at accretion shocks in large-scale structures,
 and their dynamical feedback on the clustering properties
of the thermal baryon gas. 
In a first set of simulations, we studied the large-scale
properties of CRs using fixed grid simulations (Sect.\ref{subsec:fixed}),
while in a second set we studied the details of the distribution
of CR energy in galaxy clusters re-simulated at high spatial
resolution with AMR (Sect.\ref{subsec:amr}).

\subsection{Runs with fixed resolution}
\label{subsec:fixed}
 
In a first set of cosmological runs we simulate the evolution of 
a box with a co-moving side of 80 Mpc, in a  "concordance" cosmological
$\Lambda$CDM model,  with $\Omega_{\rm dm}=0.226$, $\Omega_{\rm b}=0.044$ and $\Omega_{\rm \Lambda}=0.73$.
The normalization of the primordial index of density fluctuations was set to $\sigma_{8}=0.8$ and $h=0.72$. 
The initial redshift of the simulations is $z_{\rm in}=30$.
All runs are non-radiative, and neglect all non-thermal phenomena 
other than CR physics, such as energy release from 
stars and AGN or magnetic fields.
When re-ionization in the early Universe is modelled, we follow 
the run-time recipe introduced in \citet{va10kp}, where a 
simple prescription for a temperature floor in the simulation is adopted in the redshift range $7 \geq z \geq 2$ in order to mimic the re-ionization, as in \citet{hm96}. 
Despite the moderate resolution achieved in these runs ($\Delta x \geq  122$ kpc/h), they allow us to test the role of the parameters of our models. 
The summary of the 15 runs at fixed grid resolution is listed in Tab. \ref{tab:tab1}.

Our fiducial  model for simulating the evolution of CR energy
in our {\it {\small ENZO}} cosmological runs is 
Run3 (and Run3\_l, Run3\_m and Run3\_h at other resolutions, see below). In this model we adopt the full set of 
CR-modules of our implementation (injection at shocks, spatial advection, pressure feedback and the reduced thermalization). The injection of CRs at shocks starts from  $z_{\rm cr}=6$, and for $\rho/\rho_{\rm cr,b} \geq 0.1$ (where
$\rho_{\rm cr,b}$ is the critical baryon density of the Universe). 
The acceleration efficiency depends on $M$, as in \citet{kj07}.

In other runs we varied various parameters in order to study:

\begin{itemize}

\item the dynamical role of CRs in the evolution of large-scale structure with a simple non-radiative run (Run1). In this run,
we inject and passively advect CRs in the simulated volume, but CRs do not couple 
to the baryoicn gas (therefore they do not exert pressure  nor do they cause reduced thermalization in the post-shock);
\item the role of the reduced thermalization efficiency, by
comparing Run3 with a run with CRs injection and pressure feedback but no reduced thermalization at shocks: Run2 (in this model the total energy at the shock is not conserved, since "new" pressure from accelerated CRs is just added to the system);
\item the role of spatial resolution and DM mass resolution by re-simulating Run3 with a coarser (Run3\_l) and with two finer
resolutions (Run3\_m and Run3\_h). For comparison, we also re-simulated the same setup of Run1 with $512^{3}$ (Run1\_h);
\item the role of re-ionization by re-simulating Run3, for the extreme assumption of no
re-ionization at all (Run3\_nore);
\item the effect of assuming different epochs for the start of the injection of CRs at shocks, by investigating 
also the cases of $z_{\rm cr}=1$ (Run3\_z1) 
and $ z_{\rm cr}=29$  (Run3\_z29);
\item the effects of assuming a different minimum gas density  for the injection of CRs at
shocks, studying the case of $\rho/\rho_{\rm cr,b}=10^{-3}$ (Run3\_d0001) and $\rho/\rho_{\rm cr,b}=10$ (Run3\_d10); 
\item the effects of assuming fixed efficiencies for the acceleration
of CR at all shocks: we investigated the "toy" models where the fixed efficiencies of $\eta=0.1$ (Run3\_eta01) and $\eta=0.01$ (Run3\_eta001) are 
adopted at all shocks (with no dependence on $M$);
\item the differences caused by computing the Mach number based on the
total pressure (Run3\_mcr) instead of gas pressure, and applying
the \citet{kj07} efficiency.  
\end{itemize}

\subsubsection{Large-scale distributions of thermal gas and cosmic rays}
\label{subsubsec:distributions}

We first test the CRl modules by comparing the large-scale distributions of
Mach number, gas density, gas temperature and CRs pressure in Run1, Run2 and Run3 at z=0.5 (Fig.\ref{fig:lls_scaling}).  
The presence of CRs has no strong dynamical effect in the evolution of thermal gas on these large scales. 
Some differences are found at large over-densities (where CR feedback
reduces by a $\sim 5-10$ percent the largest over-densities in the
box), and in the temperature distributions inside accretion shocks, $T > 10^{5}$ K. 
Inside large-scale structures, Run3 presents a small deficit of gas
temperature compared to the standard simulation of Run1, due to
the fact that part of the thermal energy at accretion shocks is
channelled into energy of CRs. On the other
hand, Run2 shows a significant excess of temperatures 
within the same regions compared to Run1. The obvious explanation for this is that
the total energy at shocks is not conserved in this run, but new energy is added to the system 
when new CRs are injected. This makes outer accretion shocks stronger over time in Run2, and therefore more 
efficient in thermalizing the medium.

An important proxy to understand the dynamical role of accelerated CRs is the pressure ratio between CRs and thermal gas,
$P_{\rm cr}/P_{\rm g}$, as a function of cosmic over-density (last panel of Fig.\ref{fig:lls_scaling}. 
The trend of this ratio is broadly similar in all runs, with a 
maximum at the over-densities close to the critical one. In our fiducial model (Run3) this maximum is 
$P_{\rm cr}/P_{\rm g} \sim 0.2$, and smoothly declines to $P_{\rm cr}/P_{\rm g} \sim 0.04-0.08$ for 
$\rho/\rho_{\rm cr,b} \approx 100$. This ratio is larger if the reduced thermalization at shocks is not included ($P_{\rm cr}/P_{\rm g} \sim 0.5$ at the maximum in Run2) and even larger if the accelerated CRs are just passively advected in the simulation (Run1). 

\bigskip

We test the effect of spatial resolution and DM mass
resolution by re-simulating the same setup of Run3 with at a smaller (Run3\_l, using $64^{3}$ cells and $m_{\rm dm}=4.0 \cdot 10^{10} M_{\odot}/h$) and at two
larger resolutions (Run3\_m with $256^{3}$ and  $m_{\rm dm}=5.1 \cdot 10^{9} M_{\odot}/h$, and Run3\_h with $512^{3}$ and $m_{\rm dm}=8 \cdot 10^{7} M_{\odot}/h$). The final maps of CR pressure for the most massive galaxy
cluster in the box is shown in Fig.\ref{fig:lls3}, and suggests that the CR pressure at the largest scales in the simulation are
very similarly reconstructed at all resolutions. The increase in resolution, however, produces a more sub-structured
distribution of CR pressure inside clusters.
As expected from previous studies \citep[][]{ry03,va09shocks,va11comparison}, the increase in the spatial resolution leads to a 
sharper reconstruction of shock waves, and to a lowering of the Mach number of strong accretion shocks. When averaged over
 the whole cosmological volume,  very good convergence
in the volume distribution of simulated large-scale shocks is achieved at the resolution of $\Delta x \sim 200$ kpc/h, corresponding to the resolution achieved in Run3 (Fig.\ref{fig:lls_scaling}). 
The average pressure ratio $P_{\rm cr}/P_{\rm g}$ shows a very regular behaviour across resolutions, and a nearly converged trend with over-density for $\rho/\rho_{\rm cr,b} > 1-10$. 
At the maximum resolution here ($\Delta x = 112$ kpc/h)
the average pressure ratio at the scale of the innermost cluster regions is $P_{\rm cr}/P_{\rm g} \sim 0.05$. In
Sec.\ref{subsec:amr} we will show that this trend is confirmed even at larger resolution, using adaptive mesh
refinement for a subset of massive galaxy clusters.

\bigskip

We finally tested the various assumptions related to the injection of CRs by modifying the setup of Run3. For instance, a suitable
 minimum magnetic field is required for DSA to work, and therefore it is reasonable that the acceleration 
of CRs takes place only at over-densities large enough to have had the opportunity to generate sufficiently high magnetic
fields. 
Assuming a minimum over-density for the injection of CRs in the
simulation also reduces the number of un-necessary computations.

Figure \ref{fig:lls_scaling_eff} summarizes the differences in the final
distribution of $P_{\rm cr}/P_{\rm g}$ as a function of gas over-density
in the simulated volume, for all runs with varied parameters of CR
acceleration but identical spatial and DM mass resolution, at $z=1$ and $z=0$.  

The differences are somewhat larger at $z=1$ (top panel) and become
less significant for large-scale structures at $z=0$ (bottom). At all epochs, the maximum
of the pressure ratios is always found at $\rho \sim \rho_{\rm cr,b}$, meaning that
the accretion regions of large-scale structures are the locations where CRs can have the largest dynamical role.
As seen above, the pressure ratio inside large-scale structures is usually much smaller, $P_{\rm cr}/P_{\rm g} \sim 0.1$ for 
$\rho/\rho_{\rm cr,b}\geq 10^{2}$, in all tested models.

When the acceleration efficiency depends on the Mach number, we  find that the pressure ratio for all over-densities $\rho/\rho_{\rm cr,b}\geq 1$ is almost independent of the assumptions 
(on the initial epoch, minimum over-density for injection, re-ionization, etc.). 
Differences of one order of magnitude in the pressure ratio between the different re-simulations can be found only 
for densities below the critical one. There the effects of re-ionization and of the assumed first epoch of
DSA are still present at $z=0$.

\bigskip

In the runs using a fixed acceleration efficiency the two
trends for $\rho/\rho_{\rm cr,b}>1$ are rescaled version of the same
profile. In particular the model with the fixed efficiency of $\eta=0.1$ lies just below our fiducial
model with the efficiency of \citet{kj07} for $\eta(M)$. This suggests that at $z=0$
the bulk of the injection of energy in CRs happens with an average 
Mach number of $M \sim 2-3$ (typical of internal merger shocks), which are 
characterized by an acceleration efficiency of $\eta \sim 0.1-0.2$ (see also Fig.\ref{fig:eta}). This is also in line with our previous estimates based on post-processing the time evolution of the energy processes at cosmological shocks in {\it {\small ENZO}} runs \citep{va09shocks}.

Given the present large theoretical uncertainties conditions of DSA
in the early universe, and in presence of small gas density and very
weak magnetic fields, it is reassuring that such uncertainties
contribute only modestly to the final distribution of CR energy
inside all structures, for $z<1$. In essence, we can assume that
the average distribution of CR energy inside large-scale structures is 
almost totally dominated by the dynamics of accretion shocks and merger
shocks in the hot and dense baryon plasma, within the last $\sim 8$ Gyr.

This ensures that, at least to first approximation and limited to regions
of radii $\sim 2-3 R_{\rm vir}$ around galaxy clusters, the statistics
provided by our method for CRs is very robust, once an acceleration scenario
for CRs is specified.

\begin{figure*}
\includegraphics[width=0.95\textwidth]{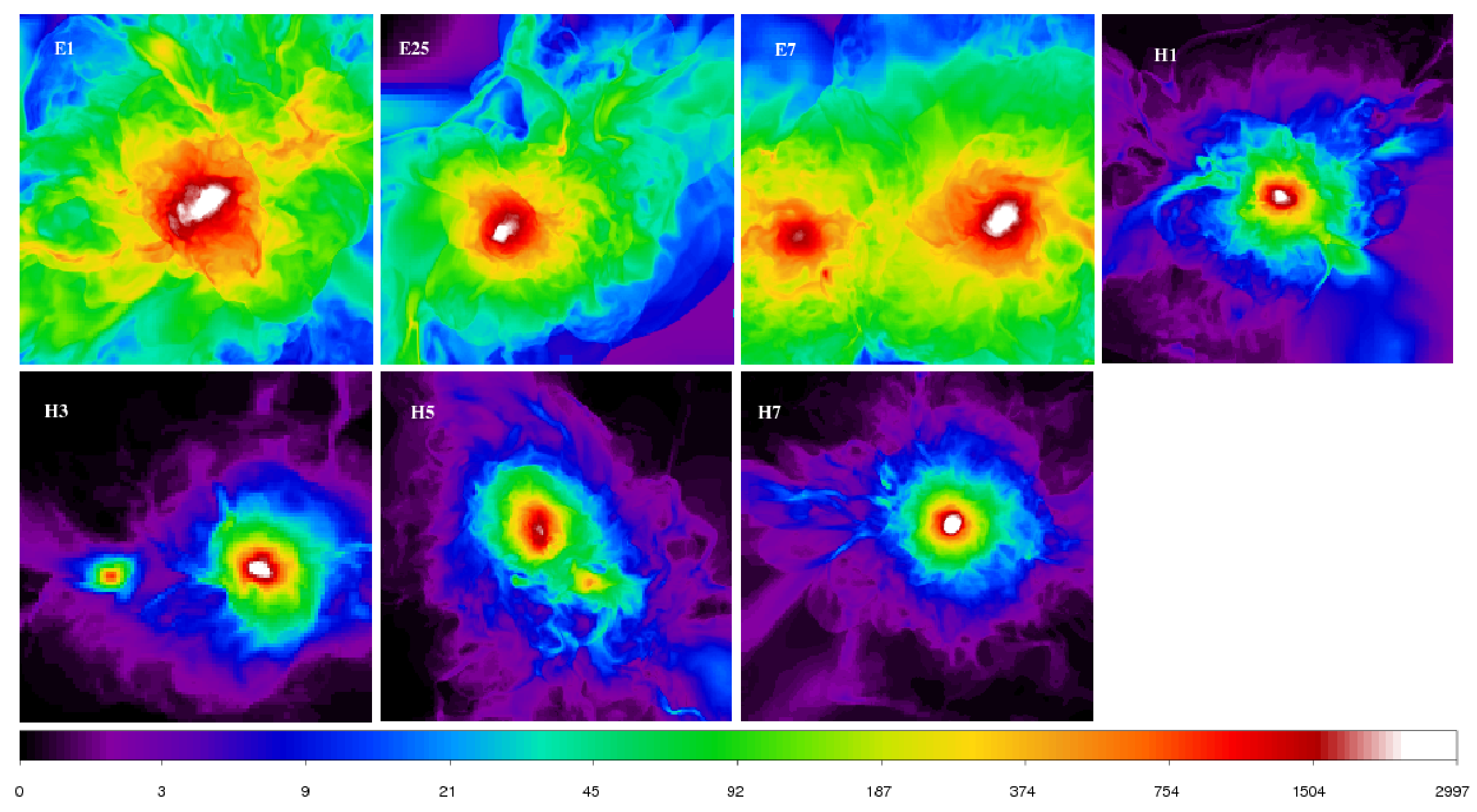}
\includegraphics[width=0.95\textwidth]{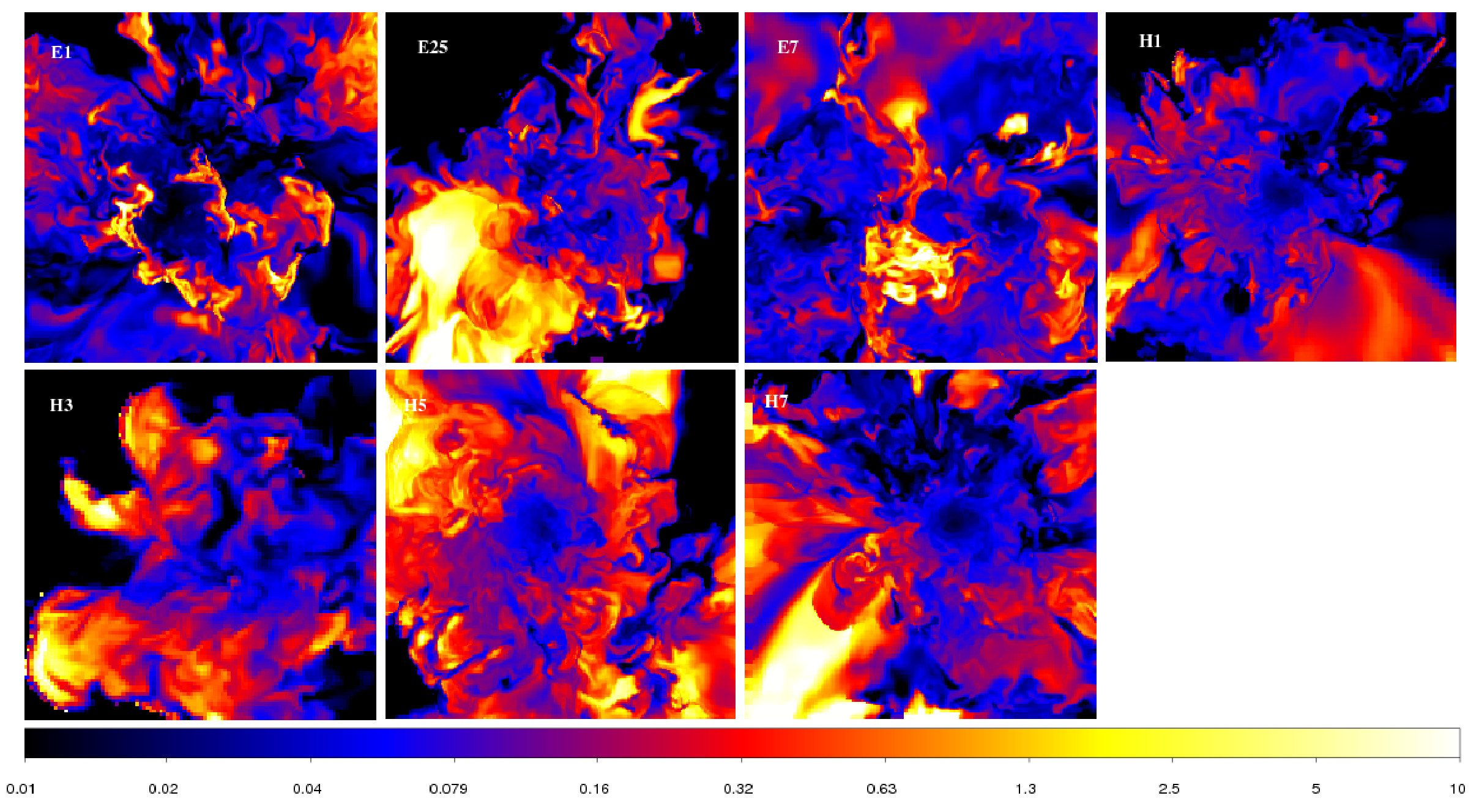}
\caption{Top panels: maps of gas density for slices of depth 25 kpc/h through
the centre of re-simulated clusters of various masses at $z=0$ (color panel in $\rho/\rho_{\rm cr,b}$). Bottom panels: maps of CRs to gas pressure ratio for the same regions. Each panel
has a side of $\approx 10$ Mpc.}
\label{fig:amr_all}
\end{figure*}

\begin{figure*}
\includegraphics[width=0.95\textwidth]{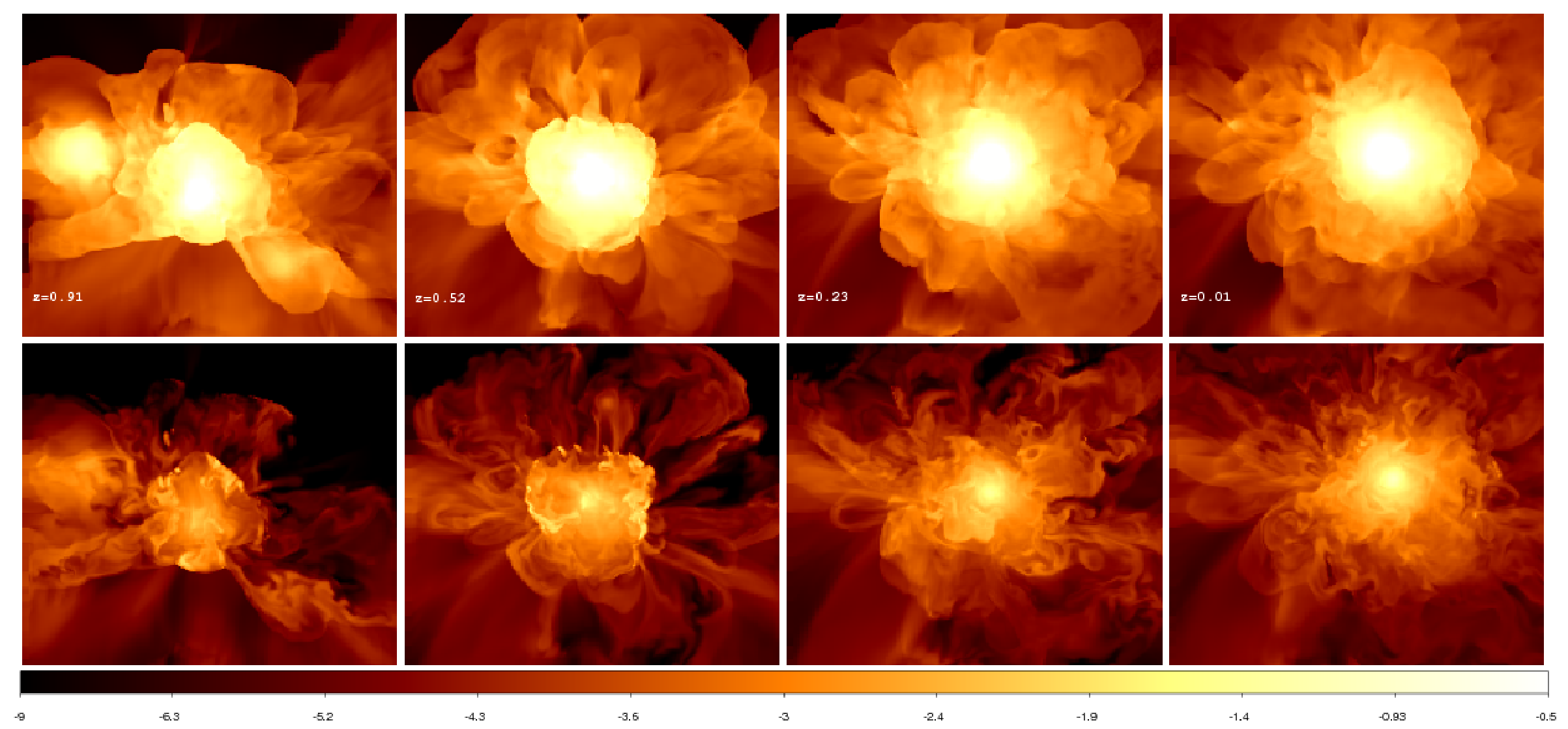}
\includegraphics[width=0.95\textwidth]{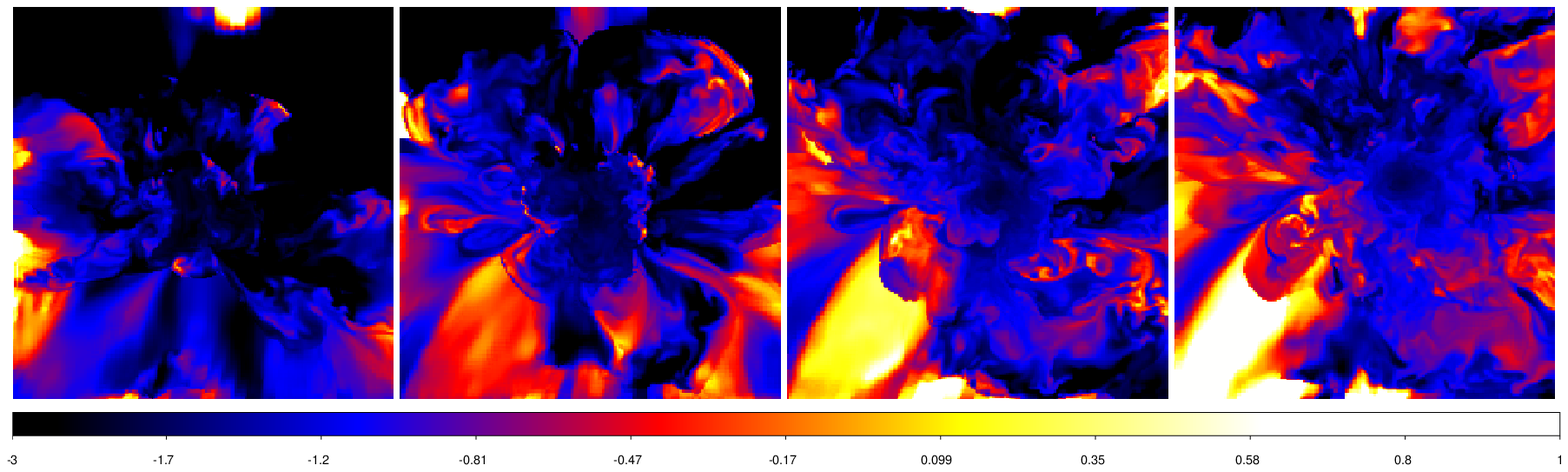}
\caption{Maps of gas pressure (first row, shown is ${\rm \log_{\rm 10}}(P_{\rm g})$), CR pressure (central row, shown is ${\rm \log_{\rm 10}}(P_{\rm cr})$) and ratio of the two (last row. $\log_{\rm 10}(P_{\rm cr}/P_{\rm g})$) for the major merger cluster H7 at four epochs. Each slice has depth 25 kpc/h and side 10 Mpc/h.}
\label{fig:amr1}
\end{figure*}

\subsubsection{Scaling relations for clusters}
\label{subsubsec:clusters}

We  first studied the global effects of CR dynamical feedback on a statistical sample of galaxy clusters, extracted
from the two most resolved runs at fixed resolution (Run1\_h and Run3\_h).

The spatial resolution of 112 kpc/h (which also represents
the softening for the computation of the gravitational force) 
is suitable to study the outer accretion regions of galaxy 
clusters and groups, but not to the resolve in detail the
mixing in the innermost cluster regions, and the
evolution of cluster cores.
For this reason, we consider the sample of clusters useful to study the effect of 
CR dynamics in the region inside $R_{\rm 200}$ (where
$R_{\rm 200}$ is the radius enclosing an average density 
of 200 times the cosmic critical density). Clusters
were extracted with a halo finder based on spherical over-density \citep[as in][]{gpm98b}.
A more detailed study of the distributions of CRs inside
clusters, at a higher resolution, is presented in Sec.\ref{subsec:amr}.

\bigskip

Our volume of $(80 {\rm Mpc})^{3}$ contains $\sim 50$ halos
with a total mass larger than $10^{13} M_{\odot}$ at $z=0$ (8 with $M>10^{14} M_{\odot}$). The panels in Figure \ref{fig:maps_projected} give the visual impression of the 
number of halos in the simulated volume, and the large-scale
distributions of gas and CR pressure associated with them. 
 Since the injection of CRs occurs only for a large enough overdensity ($\rho \geq \ 0.1\rho_{\rm cr,b}$), the
3--D distribution of CR pressure has a much lower volume filling factor than that of gas pressure.

The comparison of the $T_{\rm 200}$ vs $M_{\rm 200}$ relation and of $L_{\rm X,200}$ vs $M_{\rm 200}$ (i.e. average temperature, total mass and
bolometric X-ray luminosity inside $R_{\rm 200}$) for the most massive clusters of both runs is shown in the top panels of Fig.\ref{fig:scaling}. 

Given the limitations of physical processes included in our runs
(no radiative cooling, star formation and feedback from supernovae and AGN), the comparison of these two runs
is helpful to highlight the effects of simulated
CRs in clusters, while they do not have a strong prediction
power on observed scaling relations.

In the two runs the self-similarity
of cluster relations is preserved, but the dynamical feedback
of CR energy injected at shocks modifies the normalization 
of both scaling by $\sim 20$ percent for ($M_{\rm 200}$,$T_{\rm 200}$) and for ($M_{\rm 200}$, $L_{\rm X,200}$).
As will be shown in the following (Sec.\ref{subsec:amr}) this is a result
of a slightly modified distribution of gas matter density inside halos.
The last panel of Fig.\ref{fig:scaling} shows the pressure
ratio between CRs and gas within $R_{\rm 200}$ for all objects of
Run3\_h. The majority of clusters has a ratio of the order of $P_{\rm cr}/P_{\rm g} \sim 0.1$, with an almost constant average ratio across
one order of magnitude in mass (even though the distribution has a quite
large scatter, and no simple trend with total mass can be detected). 

\begin{figure*}
\includegraphics[width=0.95\textwidth,height=0.45\textwidth]{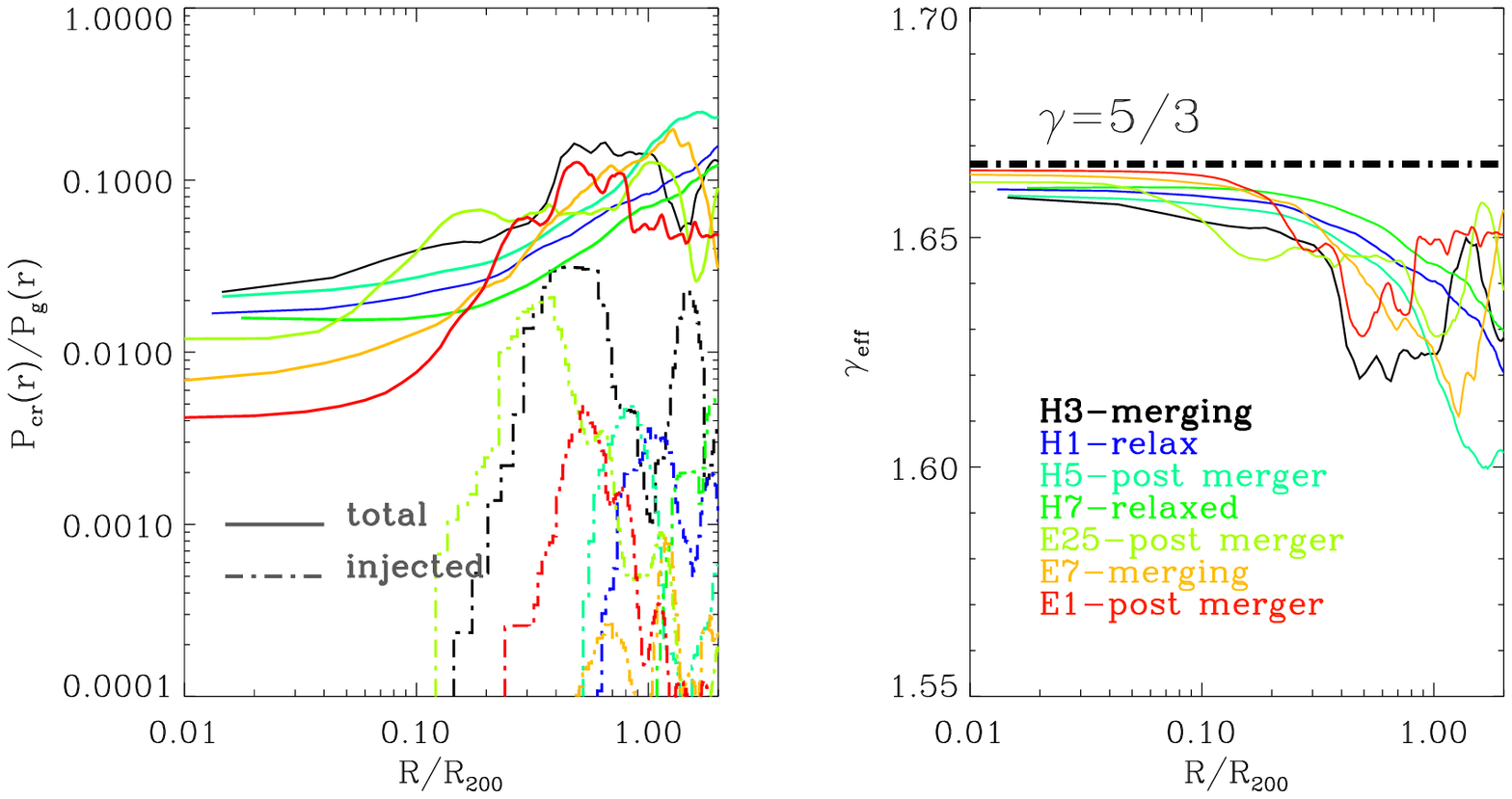}
\caption{Left panel:radial profiles of average CRs to gas pressure ratio for the sample of re-simulated clusters with AMR at $z=0$ (solid lines), and pressure ratio only for
the CRs injected during the last $\Delta t \sim 0.1$ Gyr (long dashed lines).   
Right panel: average effective (pressure weighted) adiabatic index for the same clusters at $z=0$.}
\label{fig:prof_ratio}
\end{figure*}

\begin{figure}
\includegraphics[width=0.4\textwidth,height=0.42\textwidth]{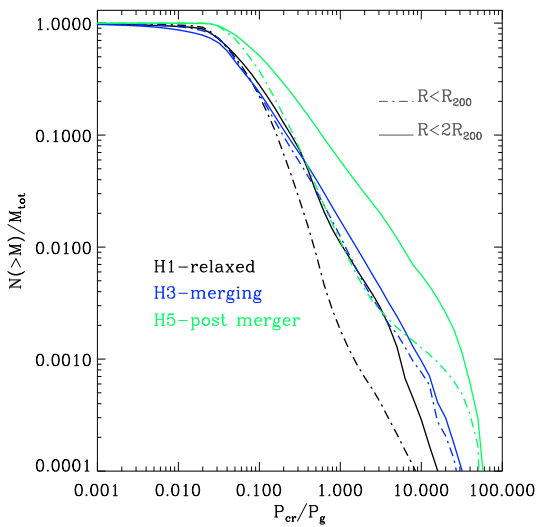}
\caption{Cumulative gas mass distribution as a function of $P_{\rm cr}/P_{\rm g}$, for clusters H1, H3 and H5 at z=0. The solid lines
are for the volume inside $\leq 2 R_{\rm 200}$, the dot-dashed lines
are for $\leq R_{\rm 200}$.}
\label{fig:prof_mass}
\end{figure}

\subsection{Adaptive mesh refinement runs}
\label{subsec:amr}

A second set of cosmological simulations was produced with adaptive mesh refinement,
achieving almost uniform high spatial resolution within cubic boxes of side length $\sim 4-6 R_{\rm 200}$ for several galaxy 
clusters in the mass range of $2 \cdot 10^{14} \leq M/M_{\odot}/h \leq 2 \cdot 10^{15}$. These runs are computationally fairly expensive (e.g. $\sim 2 \cdot 10^{4}$ hours of CPU time for the most massive clusters), and in this work we 
limit ourselves to the analysis of 7 re-simulated objects, selected from catalogues of clusters already presented in \citet{va10tracers} and \citet{va10kp}, in order to sample
different masses and dynamical state of the ICM.
 
The use of AMR is necessary here to model the turbulent motions in mergers, which may affect the properties of transport of CR energy within clusters over time. In addition, the increased resolution in capturing shocks can enhance our modelling of CR injection at internal shocks.

All objects have been re-simulated at high spatial and DM mass resolution starting from
parent simulations at lower resolution, using nested initial conditions
\citep[e.g.][]{abel98}. The mass resolution for DM inside the high resolution region where the clusters form  is $m_{\rm dm}=6.7 \cdot 10^{7} M_{\odot}$, the coarsest spatial
resolution inside this region is $200$ kpc/h while the peak one is $25$ kpc/h. In general a fraction of $\sim 60-80$ percent of the total volume inside $\leq 2 R_{\rm 200}$ is 
simulated at the highest available resolution by the end of the simulation \citep{va10kp}. In these runs we triggered mesh refinement by standard gas/DM over-density 
criteria \citep[e.g.][]{osh04} 
and velocity jumps \citep[as in][]{va09turbo}.

Table \ref{tab:tab2} summarizes the main properties of each object. The last column indicates the dynamical state of the cluster. Following \citet{va10kp} we estimated 
the dynamical state of each cluster at z=0 from its total matter accretion history
for $z<1.0$ (in the case of the lowest masses systems, $<3 \cdot 10^{14} M_{\odot}$, we consider instead $z<0.6$ for the same analysis, given their shorter dynamical
times). If a system experienced an increase of total matter by 30 percent of more
within 1 Gyr of time, we classify it as a ''post merger'' systems, or ''relaxed'' otherwise.
In the case of systems without a major merger in the past, but having an ongoing merger
process (estimated from the ratio between total kinetic and thermal energy within 
$R_{\rm 200}$) the system is classified as ''merging''.

Figure \ref{fig:amr_all} shows cuts through the centre of mass of each object of the
sample, reporting the gas density (left panels) and the pressure ratio within each cell
of the 25 kpc/h slice (right). Substantial morphological differences are present, related
to the dynamical state of the cluster: in merging systems (e.g. E7, H3) the presence
of close companions and ongoing large-scale accretions drive streams of  enriched CRs gas in the ICM, even close to the main cluster centre. In post-merger
systems (e.g. E1, H5) the regions of large $P_{\rm cr}/P_{\rm g}$ ratio are well
associated with merger shock waves, which expand out of the innermost cluster regions.

The long time evolution for one of these systems (H7) is shown 
in Figure \ref{fig:amr1} for the epochs of $z=0.91$, $z=0.52$, $z=0.23$ and $z=0.01$.
Two powerful
merger shocks are launched just after the collision of the cluster cores at $z=0.91$,  and become  powerful sites for the injection of new CRs in the ICM.  
While these shocks expand into the
outer cluster regions, their Mach number increases. Just downstream of the shocks, the local pressure ratio becomes of the order of $P_{\rm cr}/P_{\rm g} \sim 0.3$, while inside the  hot core region of the merger remnant the pressure ratio is small, $P_{\rm cr}/P_{\rm g} \sim 0.02-0.05$. This happens because internal merger shocks in their initial
phase are weak, $M \sim 2$, and thus more efficient in heating the ICM than
in injecting new CRs.  On the other hand while they expand their Mach number increases due the drop in temperature of the ICM, and the efficiency of injection
of CRs is larger.
At later times, the cluster is characterized by a pretty regular density structure for $<R_{\rm 200}$, and
regions of $P_{\rm cr}/P_{\rm g} > 0.1$ can be found only near the outer
accretion regions,  and also along filaments of cold gas ($T \sim 10^{5}$ K) in some sectors of the cluster.

\begin{figure*}
\includegraphics[width=0.9\textwidth]{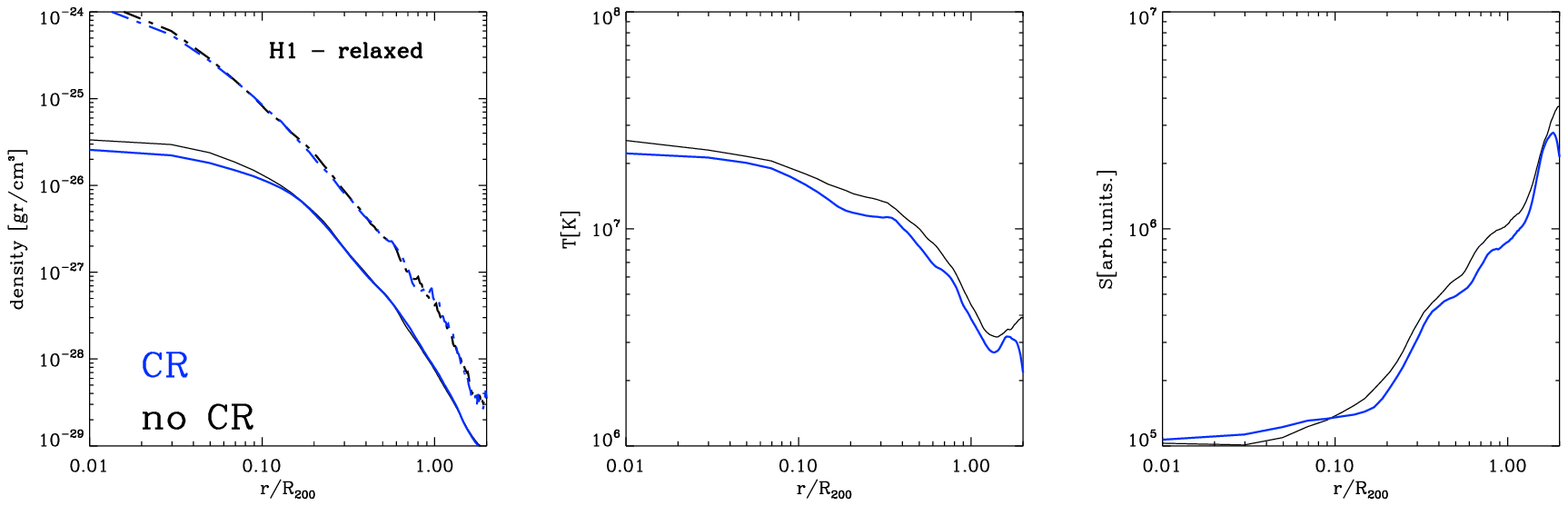}
\includegraphics[width=0.9\textwidth]{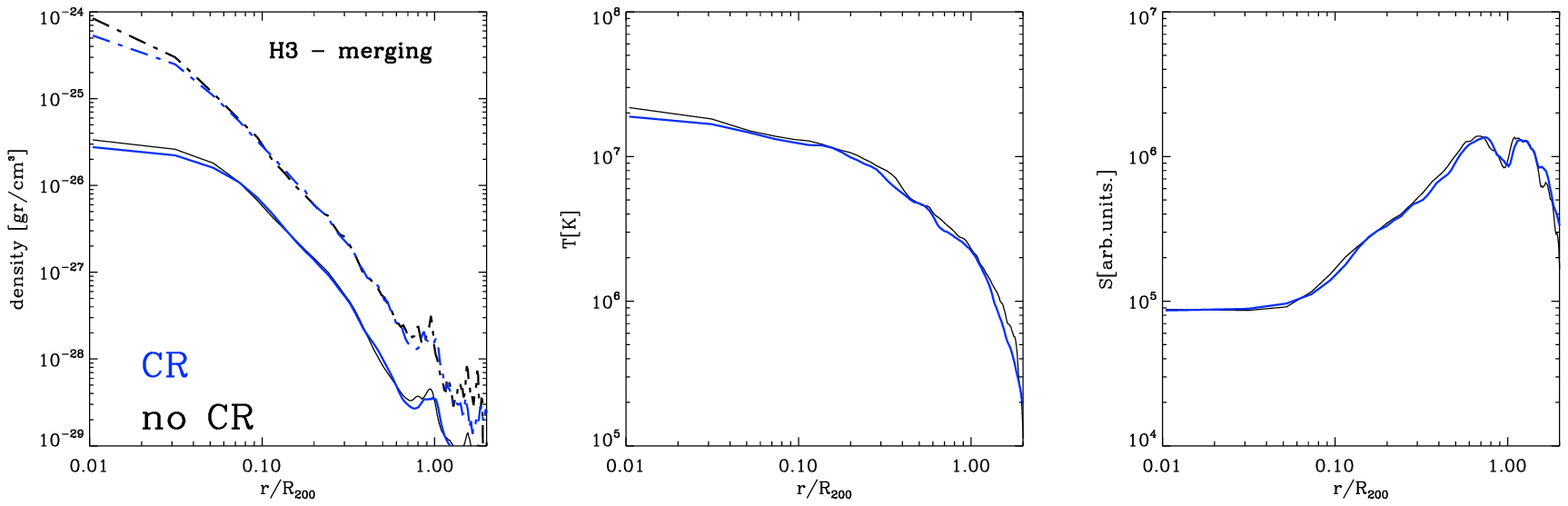}
\caption{Comparison of radial profiles of gas and DM density (left column),
gas temperature (middle) and entropy (right) for re-simulations of cluster
H1 and H3 with and without CR physics.}
\label{fig:prof_av}
\end{figure*}

\subsubsection{Radial profiles of CRs}
\label{subsubsec:amr_profiles}

Despite the different dynamical states present  in  our cluster sample, the radial profiles of the pressure ratio within each
system at $z=0$ are rather similar (left panel of Fig.\ref{fig:prof_ratio}) {\footnote {We note that characterizing the exact centre of the most perturbed systems is not trivial, see also \citet{va11turbo} for a similar discussion in the case of radial profiles of turbulent energy.}}. They show a behaviour similar to that of thermal pressure,
and are slightly flatter approaching $\sim R_{\rm 200}$. 
Within the core region of our systems, the pressure of CRs is
$\sim 0.02-0.04 P_{\rm g}$, and it increases to $\sim 0.1 P_{\rm g}$
approaching $1-2 R_{\rm 200}$. 
In the same panel, we 
also show the pressure ratio of the CRs injected at shock
waves during the last time step of each run ($\Delta t \sim 0.1$ Gyr).
Only in the case of the merging system H3,  a large amount of freshly
injected CR pressure at a peripheral shock is comparable with the thermal
pressure, while otherwise the pressure injected at 
shocks at late time is on average just a small fraction of the gas and CR 
pressure.
The right panel of Fig.\ref{fig:prof_ratio} shows the radial profile of the pressure-weighted average effective adiabatic index, $\gamma_{\rm eff}$.
Given the small ratio of CR to gas pressure in the innermost regions
of clusters, $\gamma_{\rm eff}$ is very close to the non-relativistic
value of 5/3 until the outer accretion regions are reached.
These trends are quite similar to what is reported in \citet{pf07} for non-radiative
simulations with GADGET2. However,  we notice that our profiles of CR pressure are significantly flatter, and our runs tend to concentrate
less CRs in the innermost cluster regions \citep[e.g.][]{pf07,pi10}. In Sec.\ref{sec:discussion} we will discuss the possible reasons for this. 

To highlight any dependence on the dynamical state of clusters we will focus in the following on three representative objects in the sample: 
H1 (relaxed), H3 (merging) and H5 (post-merger). These three clusters have a similar
final total mass but very different dynamical histories.

For each of them we computed  the cumulative gas mass distribution as a function
of $P_{\rm}/P_{\rm g}$ (Fig.\ref{fig:prof_mass}). In all cases,
roughly $\sim 90$ percent of the gas mass inside $R_{\rm 200}$ is characterized by a small pressure in CRs, $P_{\rm cr}/P_{\rm g} \leq 0.1$.  The post-merger
system H5 is characterized by the largest gas mass with greater
CRs pressure, followed by the merging system H3. The differences between the 
dynamical states become more prominent by considering the distributions inside $<2 R_{\rm 200}$, highlighting  the role of large-scale patterns of accretion even for
clusters with quite similar final mass. 

\bigskip

It is interesting to compare the runs with injection and feedback from CRs
to the standard non-radiative re-simulations of the same systems.
Some very general trends emerge by computing the 3--D radial profiles
of gas density, gas temperature and gas entropy (Fig.\ref{fig:prof_av}):
the innermost gas density is lower by a few percent in runs with 
CR physics. The gas entropy is correspondingly higher, while the
gas temperature is slightly lower at all radii. Also the DM
density in the innermost radial bin is decreased by a similar amount.
As shown also with fixed grid simulation (Sect.\ref{subsubsec:clusters})
this is a very general outcome of our runs,  and very similar results are 
also found for all the other re-simulated clusters of our sample.

The reason for this small differences with standard runs is non-trivial.
Starting from the early formation of cosmic structures, the injection of CRs
is efficient in the outer accretion regions, gradually leading
to a drop of the effective adiabatic index in these regions. This enhances the
gas density in the outer regions, and leaves slightly less gas matter available for the 
inflow towards the central regions of clusters. While expanding into the 
rarefying cosmic environment outside of structures, the relative pressure ratio
of CRs is further increased because of their softer equation of state (in addition, the outer thermal pressure is decreased by our modelling of the reduced thermalization efficiency at the post-shock). This process
leads overall to an enhanced total (gas+CR) pressure of the outer accretion
regions of forming structures, compared to standard runs without feedback
from CRs. At the same physical time, the structures simulated
with CR feedback have a  more extended and dense envelope of accretion
regions. This effect is similar to what 
was already found for the Zeldovich collapse test (Sect.\ref{subsec:zeld}).
In fully cosmological
simulations this effect is amplified by the fact that the expansion in 3--D 
of the outer cluster shells makes the relative increase of CR pressure more
significant. 
We further explored the reasons of the above differences between simulations with and without CR feedback
in the Appendix A.

\begin{figure*}
\includegraphics[width=0.8\textwidth]{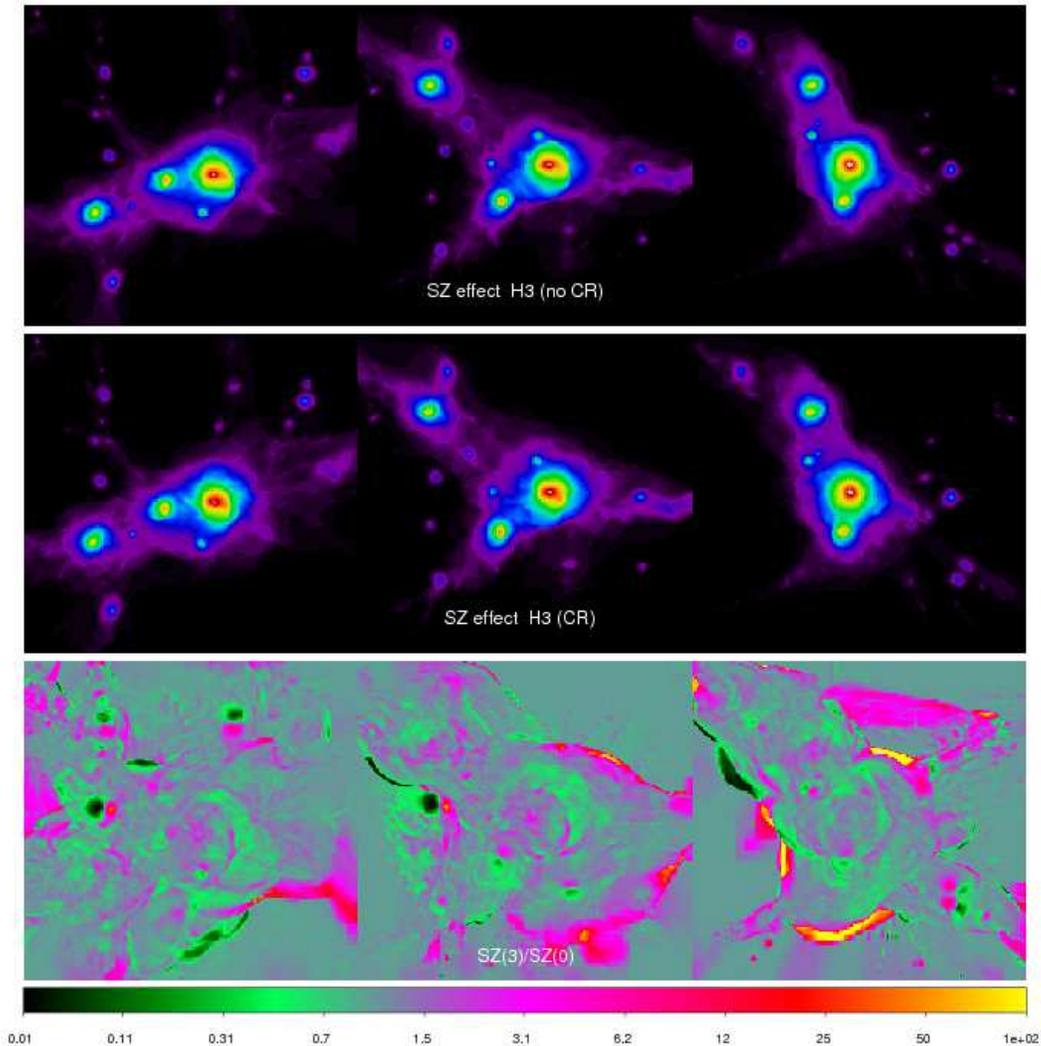}
\caption{Top panels: projected $Y$ parameter for the merging cluster H3 at z=0 in the run with CR physics, along three
projection axes. Bottom panels (and color bar): pixel by pixel ratio between the simulated maps of SZ
effect for the same cluster, obtained with (blue) and without (black) our implementation of CR physics.} 
\label{fig:sz_lx}
\end{figure*}

\subsubsection{Properties of X-ray emission and and SZ signal}
\label{subsubsec:x_sz}

Differences in the 3--D distribution of thermal gas between runs with and without CRs lead also to significant differences in X-ray emission and SZ effect \citep[e.g.][]{pf07}. 
We investigate here the projected emission maps of the simple bolometric
X-ray luminosity ($L_{\rm X} \propto \int n_{\rm e}^{2} T_{\rm e}^{1/2} dV$, where $n_{\rm e}$ is the thermal electron density and $T_{\rm e}$ is the
thermal electron temperature, which we assume equal to the gas temperature for simplicity)
and of the integrated thermal Sunyaev-Zeldovich effect ($Y \propto \int n_{\rm e} T_{\rm e} dV$) along different projections of our clusters.

In Fig.\ref{fig:sz_lx} we show the 3 maps of $Y$
for the merging cluster H3 at z=0 (simulated without CR feedback, top, and with CR feedback, central panel), and the pixel by pixel ratio obtained by dividing the maps on the top panels with the
corresponding ones from a re-simulations of H3 without CR feedback (bottom panels).

\begin{figure*}
\includegraphics[width=0.3\textwidth]{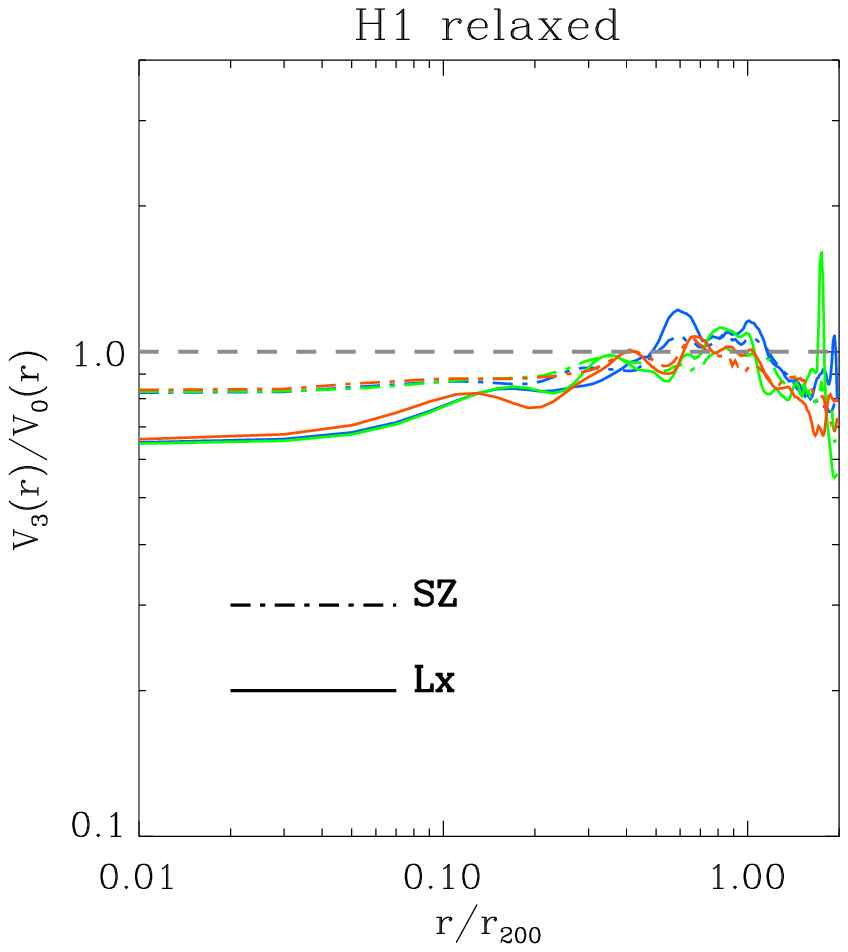}
\includegraphics[width=0.3\textwidth]{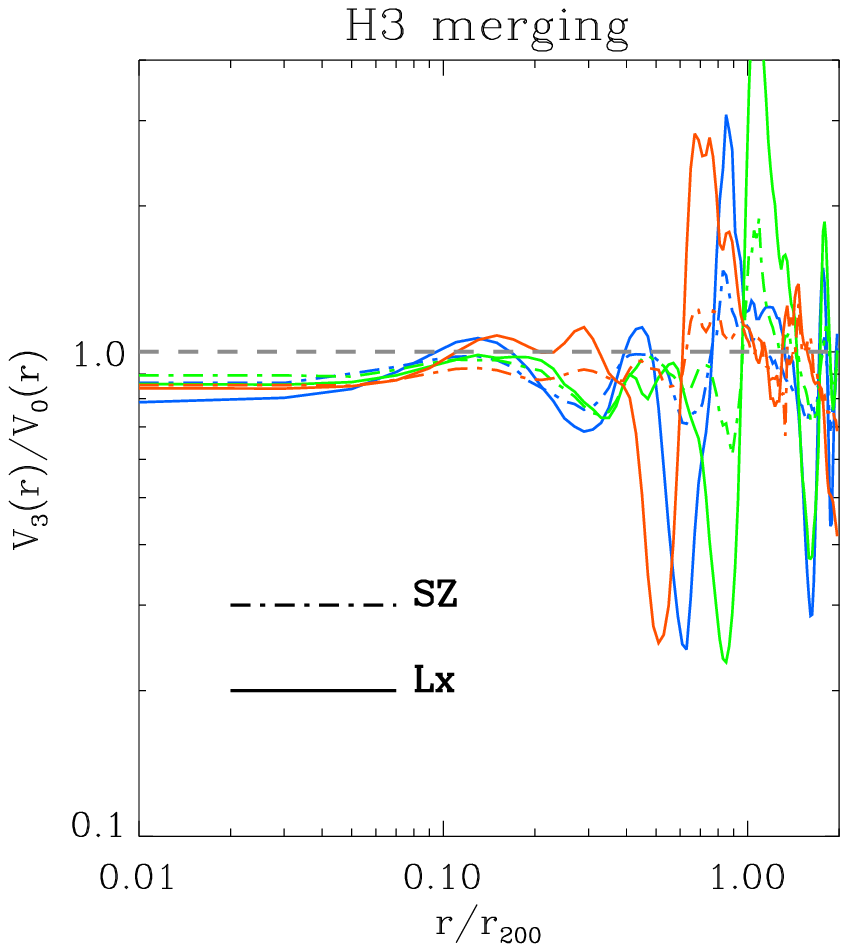}
\includegraphics[width=0.3\textwidth]{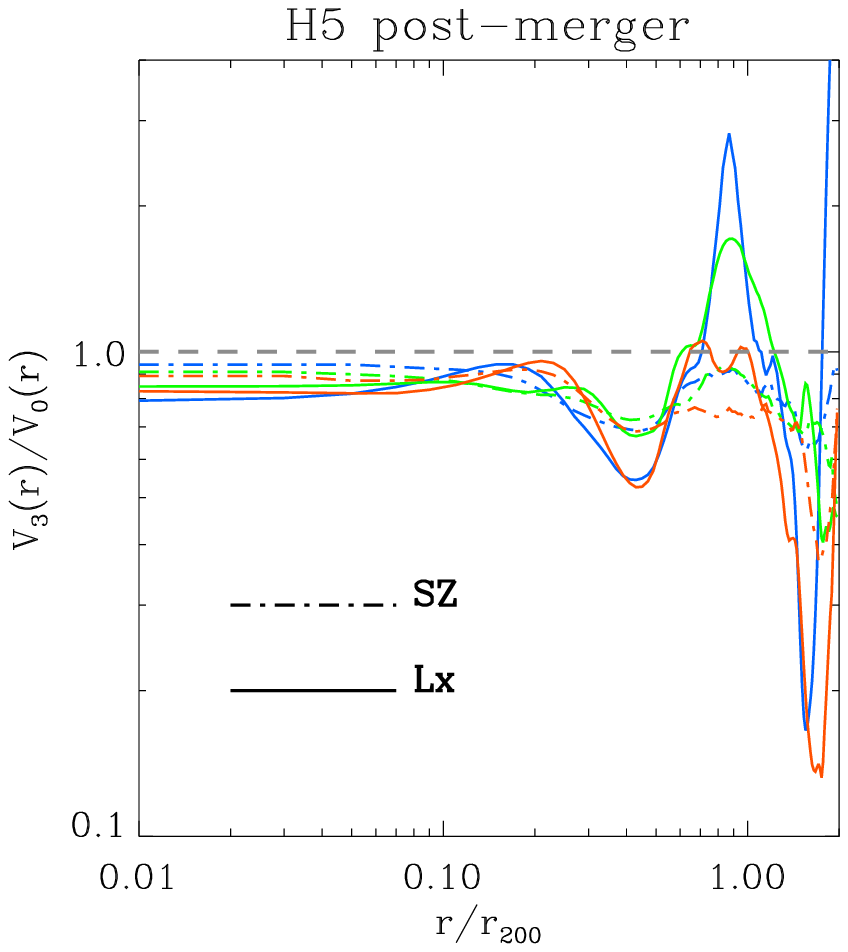}
\caption{Radial average profile of the ratio between X-ray luminosity
and SZ effect for re-simulations of clusters H1, H3 and H5 with (V3) and without (V0)
CR physics. For each cluster, the radial trend for the 3 projections are shown in different colours.}
\label{fig:prof_sz}
\end{figure*}

The large-scale morphology of the SZ effect in the two cases is extremely similar, with minor differences at small scales 
that can be partially explained by the different time-stepping
of the two runs (which affect the exact 3--D position of accreted gas
clumps in the main cluster atmosphere). The most prominent difference is related
to the outer accretion shocks, at $\sim 1-2 R_{\rm 200}$, where the run with CR feedback
produces a stronger SZ signal. The bulk of this effect is mostly
related to the physical position of the envelope of outer accretion shocks in the
run with CRs, which is farther from the cluster centre of $\sim 50-100$ kpc/h. Again, this is similar to what  observed in the Zeldovich collapse
test (Sect.\ref{subsec:zeld}): the physical effect of a mixture of CRs and thermal
gas in the outer regions produces and overall slightly accelerated expansion of the
outer shell at the same time step.
To show that this trend is general to all re-simulated clusters, we show 
in Fig.\ref{fig:prof_sz} the ratio of the radial profiles of projected maps
in clusters H1, H3 and H5, considering both X-ray emission (solid lines) and
SZ signal (dashed). 
In order to limit the very localized effect of gas clumps with small differences
in position we post-processed our simulated maps using a filtering
technique similar to  \citet{ro06}, already applied to our clusters in \citet{va11scatter}.
With this technique, the 1 percent X-ray brightest pixels at each radial bin are
excised from the raw maps, and replaced with the average value at the same radius, thus highlighting
the large-scale behaviour of the smooth gas component only.

All clusters present a similar deficit of $L_{\rm X}$ and $Y$ for $r<0.5 R_{\rm 200}$, of the order of $\sim 10-30$ percent in the case
of X-ray emission and of  $\sim 5-10$ percent in thermal SZ. 
 In the accretion regions close to $\sim R_{\rm 200}$, clusters simulated with CR feedback tend to have enhanced
$L_{\rm X}$ emission and $Y$, thus producing slightly flatter profiles compared to the runs without CRs . However, a precise quantitative estimate is more
difficult, due to the systematically larger radius of accretion shocks.
Taking the most relaxed cluster of our sample (H1) as a reference, we find that 
$L_{\rm X}$ and SZ effect increase at least by $\sim 20-30$ percent. In the case of merging clusters the maximal
difference can be as large as a factor $\sim 3-4$ in some projections.

\bigskip

Combined with the findings of the scaling for ($T_{\rm 200}$, $M_{\rm 200}$)
and ($L_{\rm X,200 }$,$M_{\rm 200}$) for the sample of clusters simulated with
fixed mesh resolution (Sect.\ref{subsubsec:clusters}), these results suggest
that the injection and pressure feedback of CRs in large-scale structures can be responsible for small 
but measurable
effects. Even if the inclusion of additional physical processes in our runs
can somewhat alter the small-scale distribution of CRs in the innermost cluster
regions (given that cooling, star formation, AGN and magnetic field should
be most important for $\sim 0.1 R_{\rm 200}$), our findings for the large
scales are more robust against the inclusion of these processes, given
that heating from shock waves is expected to be the most physically relevant
process at $\sim R_{\rm 200}$.  Interestingly, a recent analysis of a large set of clusters 
observed with {\it ROSAT} out to $\sim R_{\rm 200}$ presents the evidence for slightly flatter
outer profiles of gas density and $L_{\rm X}$, compared to the expectations of simple non-radiative simulated
cluster runs \citep[][]{eckert11}. Even if several explanation are possible for this, such
as gas clumping, the departure from simple non-radiative profiles is of the same order as that caused  by the presence of CRs.

Hence, the effect of CRs may thus have intriguing consequences in the physical interpretation
of recent observation of galaxy clusters close to  $\sim R_{\rm 200}$
\citep[e.g.][]{bau09,geo09,si11,ur11,eck11}, as well as in the use of cluster 
scaling relations to perform ''high-precision'' cosmological studies
\citep[e.g.][and references therein]{ha06,pi11}.

\bigskip

The application of cosmological simulations including feedback
from CRs were performed in the recent past with a modified version of the SPH code {\it {\small GADGET2}} \citep[][]{sp05,pf07,ju08}. 
Contrary to our results, these works found that the inclusion of CRs leads to an increase of the central  
gas density (and thus the X-ray luminosity and 
the amplitude of the SZ effect) inside $\sim 0.1 R_{\rm 200}$, while leaving
the outer cluster regions essentially un-modified. This is puzzling since the 
radial profile of effective adiabatic index in non-radiative
runs of \citet{pf07} is very similar to ours (Sect.\ref{subsubsec:amr_profiles}).

Understanding the reason of this difference is not trivial, given the different  numerical techniques and physical assumptions. 
We argue that most of the differences in the two schemes are due to the different
way in which the injection of CRs and their transport take place. In a recent
comparison of cosmological runs with {\it {\small GADGET}} and {\small ENZO} \citep{va11comparison}, we investigated
in detail how matter from accreted gas satellites is shock-heated and 
distributed within the main cluster volume.  The result is that while in {\it {\small GADGET}} matter from satellites usually retains its internal low entropy gas, delivering it
to main cluster centre ($R<0.1 R_{\rm 200}$), in {\small ENZO} the gas matter from satellites is efficiently
shock heated and deposited at much larger radii, $>0.5 R_{\rm 200}$ \citep[see also][for similar results with AMR runs]{va11entropy}. 
Very similar conclusions have been  recently presented by \citet{si11}, who compared the results on accreted cluster satellites in {\it {\small GADGET}} and in {\it {\small AREPO}}.
In the same work, we also showed that while the total energy
flux of shock waves in the cosmological volume is quite similar in the two codes, the 
average properties of external accretion shocks are not: shocks are sharper, more regular in shape and stronger in grid codes. This produces a larger entropy generation in accreted clumps in grid simulations, while much more pre-shock entropy generation is
measured in SPH runs. The effect is much more significant
than the differences due to simple resolution effects at $\sim R_{\rm 200}$in the two approaches.
Therefore, also the accretion of shock-injected CRs from accreted gas clumps
would look quite different in the two methods, once the injection of CRs
is followed in run-time. While in {\it {\small GADGET}} runs we expect that CRs from clumpy accretion
is delivered in the centre of galaxy clusters, in a grid code such as {\small ENZO}
the injection of CRs (and their dynamical feedback) mostly takes place at large
cluster radii. The net effect is that while in SPH the modification of CRs
feedback causes a larger compressibility in the centre of structures (and hence an
increase of gas density to keep pressure equilibrium), in grid codes the modification
of CRs dominates at large radii, enhancing the compressibility of the outer cluster layers already
since the early times of structure formation (see Sect.\ref{subsec:zeld} and Appendix). 
Based on our results in \citet{va11entropy} and \citet{va11comparison}, and following previous results from other groups \citep[e.g.][]{ag07,wa08,mi09}, the most likely explanation for the 
above differences is the role of artificial viscosity in the  standard SPH formulation,
and the existence of sizable pre-shock entropy generation in the standard SPH runs \citep[][]{osh04}.
Incidentally we notice also that a trend for gas density, X-ray luminosity
and gas temperature very similar to ours has been reported by \citet{rk04} for grid simulations with pressure feedback, where CRs were injected in the cluster volume by AGN-like sources.

\section{Discussion}
\label{sec:discussion}

Our simulations with an implementation of CR physics in {\small ENZO} allow us to
highlight the role of CRs injected at cosmological shocks.
Because of the numerical and physical
complexity  of CRs injection and CR feedback on the ICM plasma, 
a few strong assumptions had to be made.  

First, the acceleration efficiency at shocks is 
presently uncertain. We adopted the efficiency 
of \citet{kj07}, which is based on studies of particles acceleration in SN remnants (however in Sec.\ref{subsec:fixed} we tested also fixed efficiencies). The details of
particle acceleration in the regime of Mach numbers typical of the ICM, $M \leq 5-10$
are not yet robustly constrained by the theory, due to the difficulty of modelling the large
spatial and temporal scales involved in the diffusive acceleration at such shocks \citep[e.g.][]{kr10}. More recently, interesting studies employing particle-in-cells methods investigated
additional acceleration mechanism for particles at shocks (e.g. shock drift acceleration),
suggesting the possibilities of a different trend with Mach number \citep{ga11}.
Our treatment of CRs injection in {\small ENZO} runs will enable us to perform cosmological simulations 
with any given prescription for the functional dependence between acceleration
efficiency and background properties of the ICM.
However, the injection of CRs in the early Universe ($z \gg 2$), and in the most rarefied cosmic environment ($\rho/\rho_{\rm cr,b} \ll 1$)
is uncertain because the evolution of magnetic fields outside large-scale structures is still subject to debate
\citep[e.g.][and references therein for recent reviews]{za08,mibe11,wi11,ry11}.

Secondly, we neglect the evolution of the spectrum of accelerated particles, and
their energy losses via Coulomb and/or proton-proton collisions. The inclusion of losses may slightly decrease the  amount of CRs inside our structures. However, for the densities typical of the structures of interest here this can be safely neglected, given
that the time scales for significant energy losses of CR protons are large for typical non-cool core clusters \citep[e.g.][]{mi07,pf07}. 

Thirdly, only direct shock acceleration is considered in this work as a channel to inject
CRs in the cosmic gas. This is a safe enough assumption for most of the simulated volume,
where the turbulent \citep[e.g.][]{bl11} and the shock \citep[e.g.][]{be87} re-acceleration
of CRs should be negligible for the usually simple geometry of large-scale accretion
pattern. In the centre of clusters, however, turbulence and shocklets structures
can introduce also some significant re-acceleration of CRs. This additional channel of
acceleration is not expected to boost CR energy ratio by more than a factor $\sim $ a few over
$\sim $ Gyr time-scales. However, a future development of our method will also include these effects
(and their feedback on the gas energy) in run-time. 
The injection of CRs at supernovae and AGN is also neglected in our model. Numerical studies using {\it {\small GADGET}}
 suggested that these additional CR sources are relevant only on very small $\sim 10$ kpc scales
in proximity of the sources \citep{pf07}. Heating of the thermal gas from the Alfv\'{e}n
waves excited by streaming of CRs \citep[e.g.][]{gu08} is also an interesting physical
process which can be incorporated in future work.

The description of the ICM in the runs presented here is simplified, since we neglected radiative
cooling, and feedback from star formation and galactic activities, and magnetic fields.
Radiative cooling can have an impact on the injection of CR energy at cold lumps of gas in the
ICM \citep[e.g.][] {pf07}.  However, the handling of cooling in simulations is still problematic
since it produces strong (and un-observed) cooling flows in the inner regions of clusters, and  it is still unclear how to quench them self-consistently. 
Magnetic field, even if not dynamical important, can be relevant to the study of CR energy in
clusters because it affects the spatial propagation of CR \citep[e.g.][]{hl00,ju08}.  However, for
typical conditions in the ICM the scale-length of CR diffusion is smaller, or very close to, than our highest spatial resolution, $25$ kpc/h \citep[e.g.][and references therein]{blasi07}, making our results still valid in a statistical sense.
 
\bigskip

Within the range of assumptions listed above, our method produces a number of interesting
results with consequences on the modelling of galaxy clusters, and their
comparison with observational data.
Our runs at fixed mesh resolution show that the largest effect of CR feedback is concentrated
at the outer accretion regions of cosmic structures, at over-densities close to the
critical one (Sect.\ref{subsec:fixed}). At $\rho/\rho_{\rm cr,b} \sim 1$ we
measure $P_{\rm cr}/P_{\rm g} \sim 0.3-0.5$ on average. 
This result is very stable against
all explored modifications of numerical parameters (the pressure ratio between CRs and gas
for $\rho/\rho_{\rm cr,b} \leq 1$ is, on the other hand, significantly dependent on 
the assumed parameters). 
The use of AMR enables the investigation of the innermost region of galaxy cluster at
a higher resolution (25 kpc/h). In this work we analysed
7 galaxy clusters with different dynamical states, in the range of total masses
$2 \cdot 10^{14} M_{\odot} \leq M \leq 1.1 \cdot 10^{15} M_{\odot}$ (Sect.\ref{subsec:amr}). 
We report a very low pressure ratio in the innermost cluster regions ($\leq 0.1 R_{\rm 200}$) is very small, $P_{\rm cr}/P_{\rm g} \sim 5 \cdot 10^{-3} - 3 \cdot 10^{-2}$, increasing up to
a $\sim 0.1$ at $R_{\rm 200}$. No obvious dependence on cluster mass or dynamical state
can be found in this small sample, except for the trend that dynamically perturbed systems
display slightly larger values of $P_{\rm cr}/P_{\rm g}$, and tend to concentrate a significant
amount of CR pressure in clumps of accreted sub-halos. In all systems, the
effective adiabatic index of the mixture of gas and CRs is very close to the mono-atomic
value of $\gamma=5/3$ for  $\leq 0.5 R_{\rm 200}$. In a few perturbed systems, however,
it can be as small as $\gamma_{\rm eff} \sim 1.6$ close to  $R_{\rm 200}$.

It is important to compare the amount of CRs in our simulated clusters with
available data from observations.
A direct approach to constrain the energy
content of CR protons in galaxy clusters is the observation of 
$\gamma$-ray emission from the decay of the neutral pions due to proton-proton
collisions in the ICM. Gamma ray upper limits from {\it EGRET} observations
allow to put limits of $E_{\rm cr}/E_{\rm g} < 0.3$ in
several nearby galaxy clusters \citep{re03}. More stringent limits are derived from
deep pointed observations at energies of $>100$
GeV with Cherenkov telescopes. These limits depend on the (unknown) spectral shape of
the proton-energy distribution. For the relevant case $\delta=2$ (with $N(p) \propto 
p^{-\delta}$) the limits are $E_{\rm cr}/E_{\rm g} < 0.1$ \citep[][]{aha09,
alek10}, and they are less
stringent for steeper spectra. Recently, the  {\it FERMI} satellite greatly
improved the sensitivity of observations at
GeV energies, and the present upper limits for a large sample
of nearby galaxy clusters are $E_{\rm cr}/E_{\rm g}< 0.05$ \citep[e.g.][]{ack10,jp11}, with a poor
dependence on $\delta$. In addition to these methods, 
also the limits to the presence of diffuse Mpc-scale
radio emission in clusters can be used to
constrain secondary electrons and thus the
energy density of the primary CR protons
\citep{br07,brown11}. In this case, the limits 
depend also on the cluster magnetic
field strength and are complementary to those
obtained from $\gamma$-rays. In the relevant case of
an average magnetic field in cluster of a $\sim \mu G$,  
radio observations of clusters with no Mpc-scale radio emission
suggest that $E_{\rm cr}/E_{\rm g} \leq 0.05$, while
the limits are less stringent for smaller average magnetic fields.
These limits usually refer to innermost
$\sim$ Mpc regions of clusters, where both
the number density of thermal protons and the magnetic field are
larger. At present no tight constraints
are available for the clusters outskirts,
where the CR contribution might be larger.

Given the mean value measured inside the innermost regions of all our re-simulated
clusters (Sect.\ref{subsec:amr}),  we conclude that at present no obvious tension exists
between the estimated energy budget of CRs injected at cosmological shock waves,
at least for acceleration efficiencies in the range of those of \citet{kj07}. 
It is still possible that the inclusion of additional sources of CRs (as supernovae and AGN), as well as of
re-acceleration from turbulence and shocks may  increase the pressure of CRs close to
available upper-limits. Therefore, in the next future
a detailed comparison of $\gamma$-ray fluxes from secondary particles (possibly
with more CR acceleration mechanisms) and observations will be important to address this topic more robustly.

\bigskip

When we compare the 3--D structure of the thermal gas in these clusters
to their re-simulations which only employ standard non-radiative physics (Sec.\ref{subsubsec:amr_profiles}), we
find small systematic trends: the innermost gas density, gas pressure and entropy at z=0 are a few percent smaller if feedback from CRs is considered, while they are larger by a similar amount at $\sim R_{\rm 200}$, {\it leading to slightly
flatter radial profiles}. 
Based on our 1--D tests with the Zeldovich collapse
(Sect.\ref{subsec:zeld}) and also on specific re-simulations discussed in the Appendix A,
 we conclude that this arises from the time-integrated effect of having a lower $\gamma_{\rm eff}$
at the outer accretion regions of clusters. Accretion shocks are strong, $M \gg 10$, and their acceleration
efficiency is large. CRs are accumulated here and they become dynamically important
as the outer shells of matter of the structures expands with the growth of structures, 
because of their softer equation of state. 
This enhances the total 
pressure jump felt by infalling gas at $\sim R_{\rm 200}$ (while the gas pressure jump is slightly
decreased by the effect of having a reduced thermalization efficiency), it increases the
effective Mach number of the outer shocks, and it causes a slightly faster expansion of the
outer shells compared to standard non-radiative runs (Sec.\ref{subsec:zeld}). 
As a net effect, a few percent more gas matter is deposited in the outer regions, and 
a slightly less dense core is formed inside clusters. The reduced thermalization and 
the decreased gravitational potential in the innermost cluster regions also cause a
smaller gas pressure, compared to simulations without CR feedback, by a few percent. 

As we discussed in Sect.\ref{subsubsec:x_sz}, this feature has significant effects on the simulated properties of X-ray emission and SZ signal for our sample of clusters. 
When re-simulations with and without feedback from CR are compared, both signals are found to be decreased by 
a $10-30$ percent in the innermost cluster regions ($\leq 0.1 R_{\rm 200}$) if CRs are accelerated,  while
they are increased by factors $\sim 0.5-4$ at $R_{\rm 200}$. 
The reported trends are found to be at variance with other ones based on SPH simulations \citep[][]{pf07}, where a significant boost of X-ray and SZ signal is reported for the innermost cluster regions.
To our understanding these differences can be totally ascribed to the fundamental differences in the way the two numerical approaches model at a run-time
the accretion of gas satellites, their entropy enrichment and their stripping into
the main cluster atmosphere. This problem is related to the more fundamental ones of
artificial viscosity and pre-shock entropy generation in SPH \citep[e.g.][and references therein]{ag07,wa08,mi09,va11entropy,va11comparison,si11arepo}.

\section{Conclusions}
\label{sec:conclusions}

In this work
we presented our numerical implementation of cosmic ray injection, advection and 
feedback in cosmological simulations with the {\small ENZO} code. 
Our method incorporates a prescription for Diffusive Shock Acceleration \citep[e.g.][]{be78,dv81,kj07}, and channels at run-time energy from the thermal gas to the CR pool, thus reducing the post-shock
thermal energy flux. The main step of our algorithm are the following: a) in run-time we measure
the 3--D distribution of Mach number using a  shock finder based on pressure jumps; b) we estimate the total energy flux of  CR protons
as a function of Mach number, assuming an acceleration efficiency from theoretical models \citep[in our case, ][]{kj07}; c) we update the energy of CR energy within each shocked cell, and we advect CR energy assuming no diffusion of CRs, for a fixed adiabatic index of $\gamma_{\rm cr}=4/3$; d) we feed 
the total pressure, $P_{\rm g}+P_{\rm cr}$ (rather than the simple gas pressure, as in standard runs) into the  Riemann solver.  Therefore, the composite fluid of gas and CRs in the simulation follows an effective adiabatic index, $\gamma_{\rm eff} \leq \gamma=5/3$,  depending on the local energy ratio between CRs and the gas energy within
the cells.

We tested
our method in simple 1--D tests (Sec.\ref{sec:test1}), finding good agreement 
with analytical estimates for shock-tube tests. The role of the various parameters involved in the 
injection and advection of CRs was tested separately with shocks of different strength (Sec.\ref{subsec:tube_nocr}). 
The cosmological application of our method for CRs was studied in 1--D using the Zeldovich pancake collapse
test, where also the use of adaptive mesh refinement was tested (Sec.\ref{subsec:zeld}). 

\bigskip

We studied the injection and the evolution of CRs in large-scale structures with cosmological simulations at
fixed grid resolution (Sec.\ref{subsec:fixed}) and with adaptive mesh refinement (Sec.\ref{subsec:amr}). For fixed grid runs, we report an increase in the total CPU time of the order of $\sim 10-20$ percent compared to standard 
non-radiative runs. These low and moderate resolution runs were used to estimate the large-scale properties of 
CRs in the cosmic volume, and to test the robustness of our findings against a number of assumptions.
The level of CR energy inside cosmic structures is found to be small, $P_{\rm cr}/P_{\rm g} \leq 0.1$, with a peak at the over-density typical of outer accretion regions. We report that only the distribution of CRs outside of cosmic 
structures is strongly dependent on the details involved in the acceleration in the early cosmic epochs (and in the most rarefied environments) while the distributions of CRs are very stable for the innermost regions of clusters.

Using adaptive mesh refinement we investigated the properties of 7 galaxy clusters, with masses in the range $2.1 \cdot 10^{14} M_{\odot} \leq M \leq 1.5 \cdot 10^{15} M_{\odot}$, and different dynamical
states.
The dynamical role of accelerated CR energy is always quite small, and plays a significant dynamical role only close to $\sim R_{\rm 200}$.  In the centre of clusters instead the 
pressure of CRs is small, $P_{\rm cr}/P_{\rm g} \sim 0.02-0.05$. These values are presently consistent with the upper-limits
provided from $\gamma$-ray observations (Sec.\ref{sec:discussion}).

The effects of CRs on the overall evolution of clusters have small and
systematic effects on the 3--D distribution of the thermal baryonic gas.
In all re-simulated clusters in the innermost gas density, temperature and entropy are 
reduced by a few percent, while they are enhanced on average by the same amount at $R_{\rm 200}$ (Sec.\ref{subsec:amr}). 
This comes from the fact that CRs first modify the compressibility of 
outer accretion regions during the formation of structures, leading to an enhanced
post-shock compression and to a slightly faster expansion of the outer cluster layers
compared to standard simulations. This produces also a corresponding decrease of X-ray 
emission and of the thermal SZ signal from the inner cluster region, and an 
enhancement of a factor $\sim 0.4-4$ close to $R_{\rm 200}$, and depending of the dynamical state of the 
clusters (Sec.\ref{subsubsec:x_sz}).

 These systematic trends in galaxy clusters are at variance with results obtained with SPH for qualitatively similar treatment of CRs dynamics \citep[][]{pf07,ju08}, where enhanced gas density is found in cluster cores,
and unchanged gas distributions are measured further out. The reason of these differences is not clear. We conjecture 
that the most likely explanation lies in the different way in which transport and mixing motions are modelled in 
grid methods and in SPH, leading to a more efficient large-scale mixing of injected CRs in Eulerian runs.
This would be in line with recent comparisons performed by our group on simple non-radiative cosmological
simulations using and {\small ENZO} and  {\small GADGET} \citep[][]{va11comparison}. Performing similar comparisons
in cosmology, and with the inclusion of CRs physics, is a necessary next step to study non-thermal processes in cosmology.

To summarize, the first step for incorporating particle acceleration
at shocks in run-time in {\small ENZO}, and the physical feedback of CRs on the baryon gas is successful
for the ensemble of tests and cosmological simulations we explored. 
To our knowledge, this is the first time that such studies are successfully performed
for cosmological simulations with adaptive mesh refinement, and for realistic models of shock acceleration and reduced 
thermalization.
This model enables us to explore non-thermal phenomena in galaxy
clusters, and to take advantage 
of the capabilities of the PPM method to model turbulence and shocks
in the simulated intra cluster medium. The inclusion of more complex treatment of CR dynamics (e.g. using spectral energy bins, \citealt{mi07}), as well as the 
porting of these methods to more sophisticated and recent development of {\small ENZO}
\citep[e.g.][]{co11} are under way, and will allow us to simulate the non-thermal features of the ICM with
 unprecedented detail.

\section*{acknowledgements}
F.V. and M.B. acknowledge support
from the grant FOR1254 from the Deutsche Forschungsgemeinschaft. 
F.V. acknowledges the 
usage of computational resources under the CINECA-INAF 2008-2010 agreement, and at the at the John-Neumann Institut at the
Forschungszentrum J\"{u}lich. 
G.B. acknowledges partial financial support from PRIN-INAF 2009 and ASI-INAF I/009/10/0.
We are grateful to S. R\"{o}sswog for his help in the  1--D tests of the code, to J. Favre for his support in the use
of  {\it VISIT}. We gratefully acknowledge the {\small ENZO} development group for providing 
helpful on-line documentations and tutorials (http://lca.ucsd.edu/software/enzo/).

\bibliographystyle{mnras}
\bibliography{scienzo}

\bigskip

\appendix

\section{Investigating the reason of a lower central density}

\begin{figure*}
\includegraphics[width=0.9\textwidth,height=0.55\textwidth]{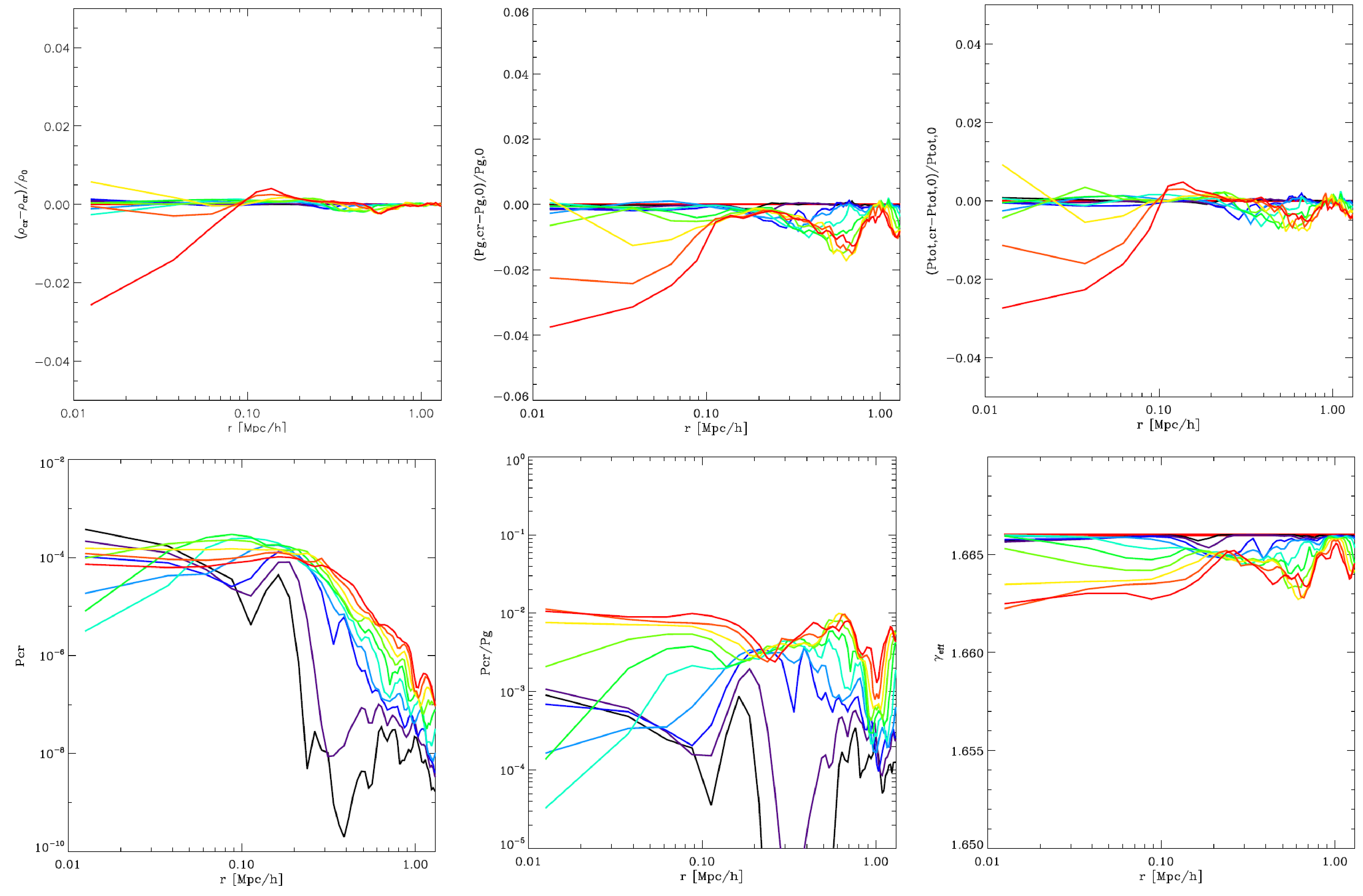}
\caption{First three panels: time evolution of the ratio of gas density, gas pressure and total pressure in a
re-simulated halos of run H3 with and without CR physics. Last three panels: time evolution of the 
CRs pressure, pressure ratio and effective adiabatic index for the same cluster. The injection
of CR has been started at z=2 in this case. The 
time sequence goes from blue to green to red colours, the time lag between each profile
is $\Delta t \sim 0.05$ Gyr.}
\label{fig:seq}
\end{figure*}

In this Section we present a numerical test designed to highlight the impact of CR physics on the
formation of galaxy clusters, at an early stage of their assembly. 
In this test, we selected a cubic sub-volume of side 10 Mpc/h around the
forming cluster H3 at z=2. 
This is a standard non-radiative run in the redshift range $30 \geq z > 2$. Starting
from $z=2$,  we evolve two re-simulations of the same volume, allowing CRs injection and feedback
in the first, and following standard non-radiative physics in the second.  
At each root-grid time step ($\Delta t \sim 0.05$ Myr) we compare the thermal properties 
of gas halos in the two runs. This highlights
the timely role of CRs in our runs, which is responsible of the significant
differences discussed in Sect.\ref{subsec:amr}.
In the first three panels of Fig.\ref{fig:seq} we show the time-sequence of profiles for the normalized difference
of gas density, gas pressure and total pressure for a massive halo in our sample ($M \sim 4 \cdot 10^{13} M_{\odot}$ at z=2). In detail, we compute
$(V_{\rm cr}(R)-V_{\rm 0}(R))/V_{\rm 0}(R)$, where $V_{\rm cr}(R)$ is the profile
of each quantity in the run with CRs feedback, and $V_{\rm 0}(R)$ is the corresponding profile in the
run with no CRs feedback. The last three panels of the same Figure report the time 
evolution of CR pressure, pressure ratio and $\gamma_{\rm eff}$
for the re-simulation with CRs feedback.

The injection of CRs within halos starts mainly from
outer accretion shocks (at $\sim 0.2$ Mpc/h in the Figure),
and at internal merger shocks at the centre
of the structure. The energy feedback of CRs is initially more
important in the outer proto-cluster regions (since the Mach number is large here), 
leading to a small and continuous decrease
of the effective adiabatic index of the mixture of gas and CRs. 
The gas pressure in the run with CRs after a few time-steps is initially reduced by a $\sim 1-2$ percent, 
while the total pressure remains almost constant. The post-shock gas density 
at the outer regions is progressively increased by the enhanced compressibility
at the shock. This in turn leads to a smaller deposition of gas matter
(reducing gas density by $\sim 1$ percent) at the outer cluster parts, compared
to the standard run. The outer shells expand into the rarefying cosmic volume, and the
smaller adiabatic index of CR energy makes it even more dynamically important as the expansion goes on. and
By the end of our test ($z \approx 1.6$) this has caused the increase of $\sim 5$ percent
in the total (gas+CR) pressure at the outer proto-cluster regions. 
We thus expect that shock waves are slightly stronger at $\sim R_{\rm 200}$ in the run with CRs, 
even at later times. This makes the injection of new CRs an even more efficient process, and maintains the difference
with standard runs even at later times, as observed in our clusters at z=0 (Sect.\ref{subsec:amr}).

 \end{document}